\documentclass[default,sn-standardnature,iicol]{sn-jnl}

\usepackage{amsmath}
\usepackage{subcaption}
\usepackage{times}
\usepackage{algorithm}
\usepackage{algpseudocode}

\jyear{2022}%

\theoremstyle{thmstyleone}%
\theoremstyle{thmstyletwo}%

\theoremstyle{thmstylethree}%

\begin{document}

\title[DER Coordination]{Coordinating Distributed Energy Resources for Reliability can Significantly Reduce Future Distribution Grid Upgrades and Peak Load}


\author*[1]{\fnm{Thomas} \sur{Navidi}}\email{tnavidi@stanford.edu}

\author*[1]{\fnm{Abbas} \sur{El Gamal}}\email{abbas@ee.stanford.edu}

\author*[1,2]{\fnm{Ram} \sur{Rajagopal}}\email{ramr@stanford.edu}

\affil*[1]{\orgdiv{Department of Electrical Engineering}, \orgname{Stanford University}, \orgaddress{\street{350 Serra Mall}, \city{Stanford}, \postcode{94305}, \state{CA}, \country{USA}}}

\affil[2]{\orgdiv{Departments of Civil \& Environmental Engineering}, \orgname{Stanford University}, \orgaddress{\street{473 Via Ortega}, \city{Stanford}, \postcode{94305}, \state{CA}, \country{USA}}}

\abstract{Current DER coordination schemes, such as demand-response and VPPs, aim to reduce electricity costs during peak demand events with no consideration of distribution grid reliability. We show that coordinating DERs for grid reliability can significantly reduce both the infrastructure upgrades needed to support future increases in DER and electrification penetrations and peak load. Specifically, using a power flow driven simulation-optimization methodology, we compare the potential reliability improvements with a perfect-foresight centralized DER controller that minimizes reliability violations to a local controller that minimizes consumer electricity cost. We find, for example, that by 2050 with local control, on average 81\% of the transformers in a distribution grid experience violations, compared to 28\% with centralized control, which simultaneously reduces peak load by 17\%. These reductions are achieved with only 5.1\% increase in electricity cost. These findings suggest that future incentives for DER adoption should include reliability coordination.}

%




\maketitle

\section*{Introduction}\label{sec:intro}


Future electric distribution grids are projected to have ever higher demands due to increased electrification of transportation, space heating, water heating, and cooking which currently use mostly fossil fuels. They will also have significantly higher penetrations of distributed energy resources (DER), most notably, roof-top solar Photovoltaic (PV) systems, stationary battery storage systems, and appliances whose consumption can be flexibly controlled, such as electric vehicle (EV) chargers, HVACs, and water heaters. While these changes bring significant benefits to consumers and to the environment, they could adversely impact distribution grid reliability, in particular, by increasing transformer overloading and voltage violations~\cite{mitnick2015changing, reliability_survey, PV_regulators_journal, EV_impacts_nature}. Combating such impacts would require costly infrastructure upgrades and / or the development of novel DER coordination schemes that jointly control DERs across a distribution grid, and electricity tariff structures and incentives~\cite{mitnick2015changing, cost_upgrades_NREL, upgrade_cost_PV, national2017enhancing, price_transactive_nature, price_dynamic_residential, price_DLMP}. To judiciously plan these future developments, there is a need to quantify the potential impacts of increased DERs and electrification on grid reliability. Such quantification, however, is challenging because it depends not only on the projected increases in DER penetration and electrification and how they are distributed across the grid, but also on the DER control and coordination schemes, electricity tariff structures, and the analytical and computational methods used. 

In this paper, we use a power flow driven simulation and optimization methodology to quantify the adverse impacts of projected increases in DER penetration and electrification on distribution grids under different DER control schemes and electricity tariff structures. Our aim is to provide answers to questions such as: How much grid infrastructure upgrading will be needed over the next three decades? How much can DER coordination help improve reliability, hence reduce such needed infrastructure upgrading? How does DER coordination impact consumer electricity cost? How much can DER coordination for grid reliability reduce peak load in the network? How do the results depend on the amount of flexible load available? How do they depend on the spread of storage across the network and on EV charger rated output power? Quantitative answers to these questions can help inform policies to achieve high future grid reliability in a cost effective manner. 
\smallskip

\begin{figure}[t]
	\centering
	\includegraphics[scale=0.635]{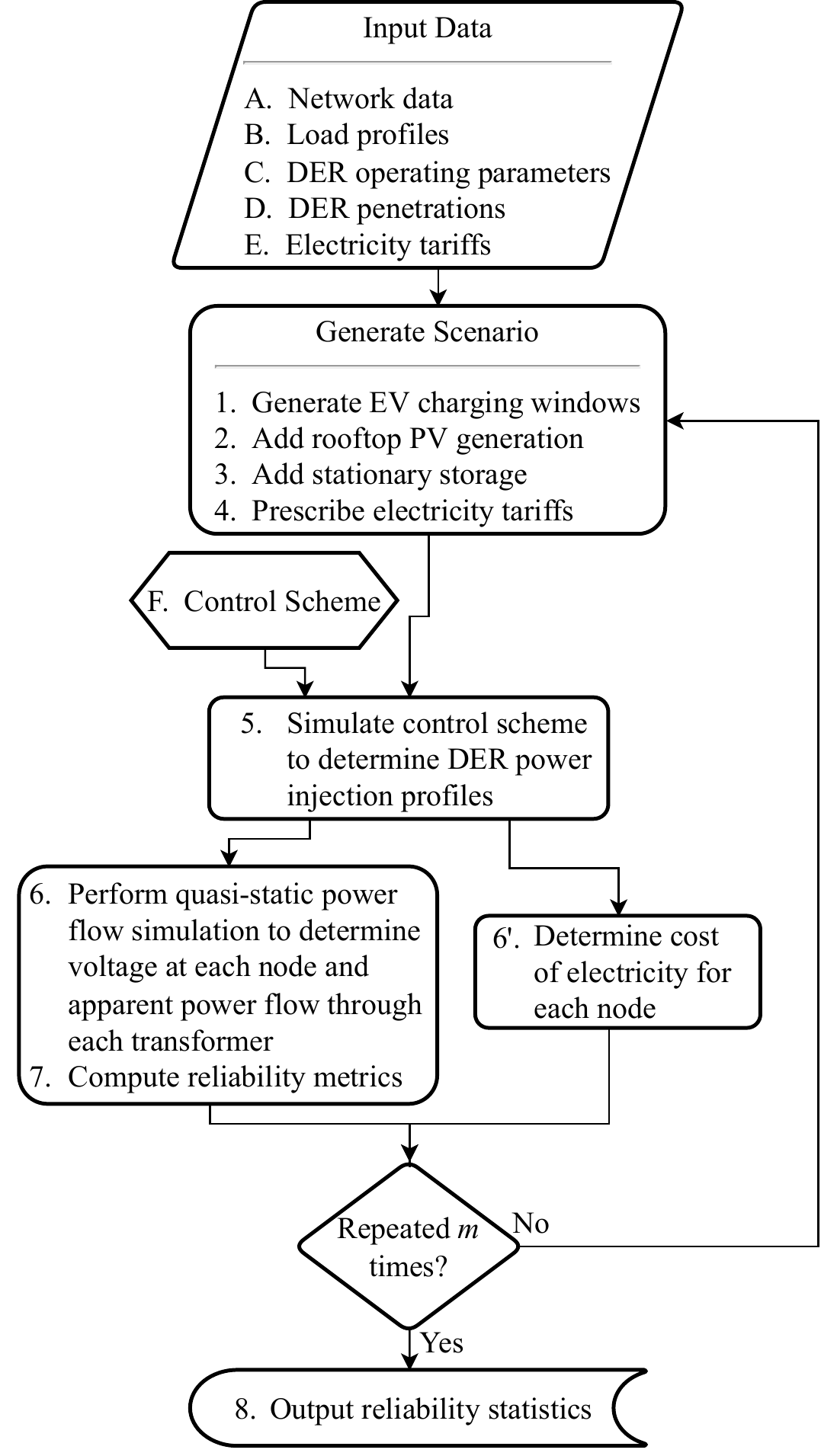}
	\vspace{4pt}
	\caption{\textbf{Block diagram of the power flow driven simulation and optimization methodology.} User specified input data are described in the parallelograms; user specified control scheme in the hexagon; and simulation operations in the rectangles. Each item in the diagram is described in detail in the Experimental procedures section. The simulation is repeated for $m$ independently generated scenarios from which reliability metrics and electricity cost statistics are computed. As will be discussed, this simulation is repeated for the local and centralized controllers, for different networks, and every 5 years between 2020 and 2050 to trace the trajectory of the impacts of DERs.}
	\label{fig:block_diagram}
\end{figure}

\noindent{\bf Simulation methodology}. A block diagram of our simulation and optimization methodology is depicted in Figure~\ref{fig:block_diagram}, which is applicable to any set of input data and control schemes. The input data comprises a 3-phase distribution network model, consumer load profiles for a chosen simulation horizon and time resolution, DER parameters and penetration levels, including the percentages of flexible load, and electricity tariffs. Given this information, a network use scenario, which specifies the placements and allocations of DERs, EV charging windows, and prescribed electricity tariffs is randomly generated. Given a DER control scheme, power injections at consumer nodes with DERs are determined for the scenario. The next steps involve computing the cost of electricity at each node and using a quasi-static power flow simulator to determine the voltage at each node and the apparent power flow through each transformer in the network, including simulation of the operation of voltage protection equipment. The metrics for transformer thermal overloading and steady-state voltage violation at each node in the network are then evaluated. This process is repeated several times and node voltage, transformer violations, and electricity cost statistics are computed. Further statistical analysis is performed to compare the results using different control schemes and across different networks.

\smallskip

\noindent{\bf Simulation setup}. We used our methodology to generate scenarios for a suite of eleven 3-phase distribution network models representing different urban, suburban, and rural areas with varying mixes of residential and commercial consumers and climates within the continental United States~\cite{grid_models_NREL, iowa, test_feeders_IEEE, EPRI_J1}; see Table~\ref{tab:network_1}.
To forecast load profiles for future years, we use the NREL dataset~\cite{restock_nrel, NREL_stock_data} to generate load profiles for the year 2018 then add electrified load profiles until we reach the projections from~\cite{NREL_EV_adoption}. The load data is organized by census code, which gives us an approximation of the population density. The EV penetration is determined using the projections in~\cite{NREL_EV_adoption} and the rooftop PV and stationary storage penetrations are obtained from the projections in~\cite{NREL_storage}. The flexible load projections are obtained from~\cite{flex_load_NREL}. We assume that all flexible load is part of the thermal loads, specifically, electric furnaces, air conditioners, and water heaters, which is consistent with projections in~\cite{flex_load_NREL}. We evaluate consumer electricity costs using PG\&E time-of-use (TOU) electricity tariffs~\cite{pge_tariffs}. In most of our simulations, we assume EV charger rated output power of 6.3 kW, but we also quantify the changes in the results of using higher powers. Using this input data and the randomized methods for sizing and distributing DERs and for generating EV charging windows as described in the Simulation methodology section, we generate a network use scenario that spans 1 year at 15min resolution. 

Since our goal is to quantify the benefit of DER coordination to grid reliability rather than evaluating any particular coordination scheme or comparing such schemes we consider two extreme controllers. The first is a local controller similar to what is in use today and whose goal is to minimize each consumer's electricity cost. At the other extreme, we consider a perfect-foresight centralized controller in which all future loads, PV generation, and EV charging windows are known two days in advance. The controller jointly optimizes all EV charging, stationary storage power injections, and flexible loads in the network to minimize the transformer overloading and voltage metrics from~\cite{hosting_capacity}, in addition to electricity cost. While this centralized controller is not implementable, it provides a lower bound on the number of violations achievable by any implementable DER coordination scheme that does not have such perfect knowledge of the future.

\smallskip
\noindent{\bf Related work}. While our simulation methodology is similar in some respects to dynamic hosting capacity analysis~\cite{capacity_analysis_survey,  capacity_analysis_dynamic_NREL, capacity_analysis_EPRI, capacity_analysis_storage_alg, capacity_analysis_storage_swiss, capacity_analysis_EVs_NREL}, it has different goals and at the same time can extend existing hosting capacity analysis in several directions. While hosting capacity is used mainly for short-term grid upgrade planning to determine the highest DER penetration at which the grid can operate satisfactorily, our methodology aims to inform future distribution grid planning and operation policies by quantifying the long term impacts of realistic scenarios of simultaneously high solar PV, EV, storage, thermal flexible loads and electrification penetrations under different DER control schemes and electricity tariff structures on grid reliability. In comparison, most existing hosting capacity analyses do not consider solar PV, EV, storage and flexible loads simultaneously and do not consider DER coordination or electricity tariffs. A more detailed review of existing hosting capacity analysis is given in the Supplement.

Previous work on simulating the impacts of DERs have addressed other types of questions than the ones posed in this paper. The study in~\cite{tariff_house_nature} investigates how increasing self-generation and storage can change residential load profiles and tariffs, and how those changes impact household expenditures. In~\cite{EV_impacts_nature}, the authors investigate the impact of increasing EV charging demand on residential load profiles and how it leads to higher peak load, hence a need for grid upgrades. Another study on the impact of EV charging~\cite{VGI_nature} focuses on the benefits of vehicle to grid coordination and how it can reduce wholesale electricity cost. The authors in~\cite{DER_LCOE_ram_nature} simulate how growing DER penetrations can reduce the levelized cost of energy (LCOE). The study in~\cite{grid_infrastructure_inequity_nature} analyzes DER hosting capacity over many distribution grids and concludes that the capacity is inequitable due to grid infrastructure limits. 

DER control today is largely performed locally at each consumer's site and aims mainly to reduce individual consumer's electricity cost with no consideration of load or solar PV forecasting and no communication or coordination with grid operators to provide any benefits to grid reliability, such as smoothing out fluctuations in supply and demand balance~\cite{solar_compete_nature}, except during infrequent demand response or virtual power plant (VPP) events. Several works have proposed methods for coordinating the operations of DERs (also referred to as Distributed Energy Resource Management System (DERMS)) across different consumer sites with the goal of reducing the adverse impacts of DERs on grid reliability~\cite{price_DLMP, capacity_analysis_storage_alg, data_opf_guo, DMPC, boyd_messaging, junjie, consensus, P2P_VPP_nature, kyle_anderson:2017, Navidi_smartgridcomm, setpoint_track_Bernstein, PacketizedEnergy, data_ML_local, horowitz_derms} or aggregating DER resources to provide grid services~\cite{Xi_coop, kyle_coop, Zhang_coop, NEC_coop}. These methods range from fully centralized~\cite{data_opf_guo, DMPC}, fully distributed~\cite{DMPC, boyd_messaging, junjie, consensus, P2P_VPP_nature, horowitz_derms}, to hierarchical~\cite{price_DLMP, capacity_analysis_storage_alg, kyle_anderson:2017, Navidi_smartgridcomm, setpoint_track_Bernstein, PacketizedEnergy, data_ML_local}. Although most of the previous work consider only solar and storage, some consider flexible loads such as EVs, water heaters, and HVAC units~\cite{PacketizedEnergy, building_carbon, building_resource, building_resource_prev, space_heating, tracking_thermal, exergy_replacement}. These works discuss the unique communication and control challenges that come with implementing any particular control scheme. In the Supplement, we provide a more detailed review of previous work on DER coordination.

\smallskip



\section*{Results with and without DER coordination}

\begin{figure*}[htpb]
\centering
\begin{tabular}{cc}
\begin{subfigure}[b]{0.5\textwidth}
\centering
\includegraphics[scale=0.6]{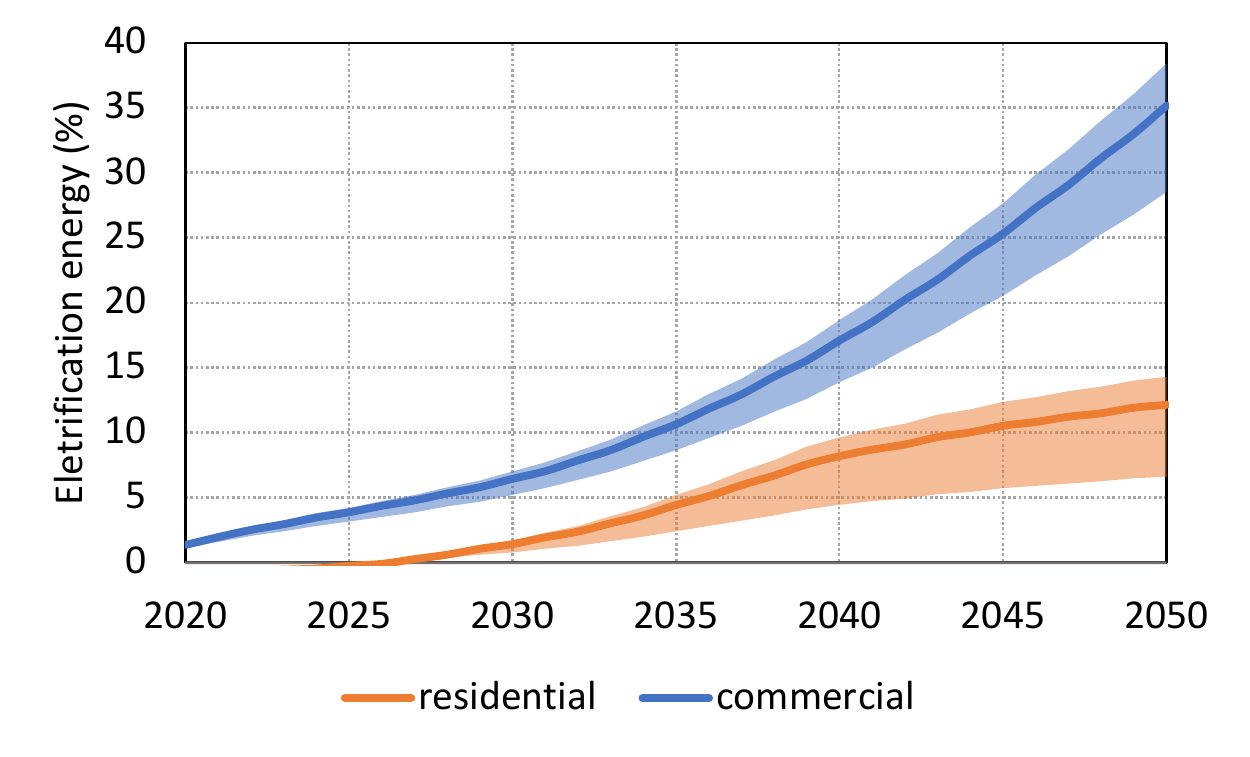}
\caption{Percentage increase in total energy due to electrification.}
	\label{fig:e_years}
\end{subfigure}
&
\begin{subfigure}[b]{0.5\textwidth}
\centering
\includegraphics[scale=0.6]{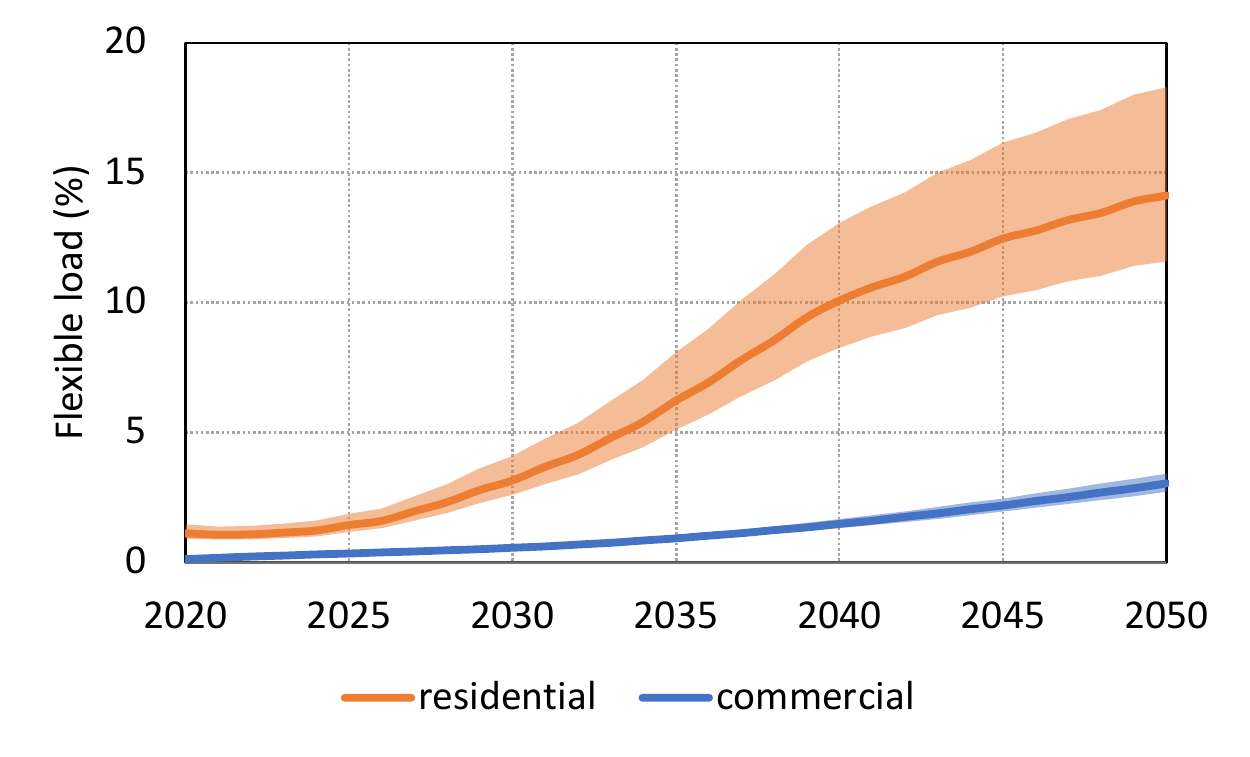}
\caption{Percentage of total energy that can be flexibly controlled.}
\label{fig:f_years}
\end{subfigure}
\\
\begin{subfigure}[b]{0.5\textwidth}
	\centering
	\includegraphics[scale=0.6]{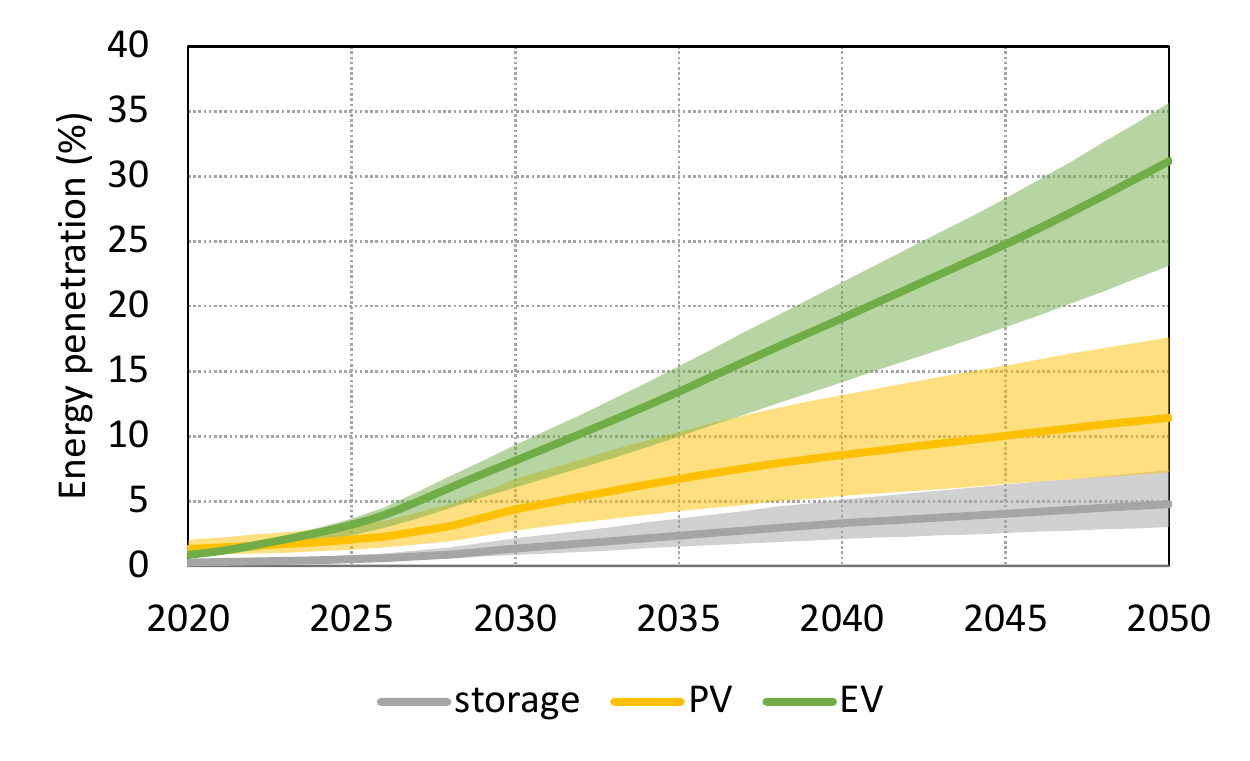}
	\caption{Percentage of energy from each DER.}
	\label{fig:pder_years}
\end{subfigure}
&
\begin{subfigure}[b]{0.5\textwidth}
	\centering
	\includegraphics[scale=0.6]{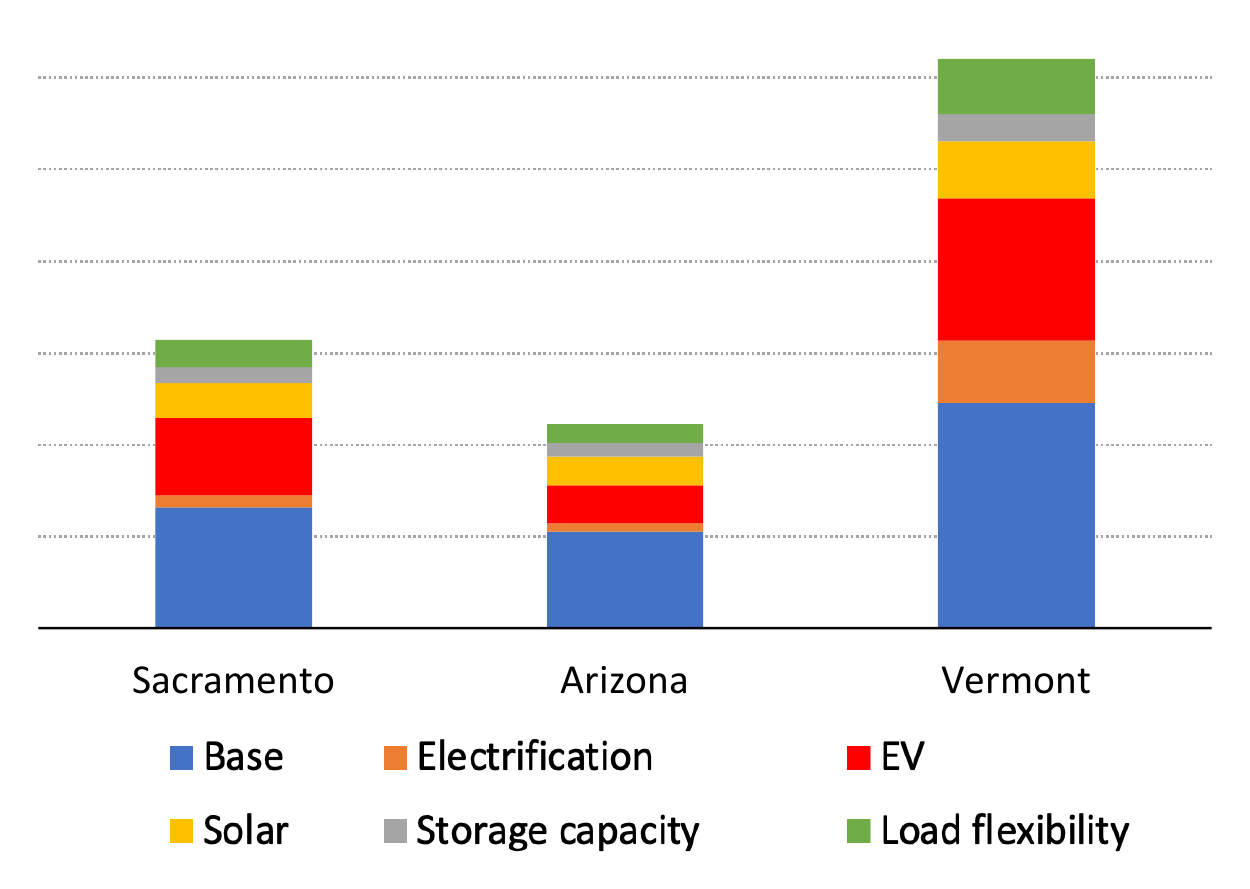}
	\caption{Examples of the energy compositions in 2050 for three networks. Each vertical tick represents 50 GWh and stacking is only meant to represent relative magnitudes.}
	\label{fig:e_example}
\end{subfigure}
\end{tabular}
\caption{\textbf{Projected means and standard errors of electrification and DER energy penetrations averaged over all networks up to 2050.} (a) The electrification energy is the percentage increase in total demands over that of the baseline 2018 demands. Note that electrification is projected to be significantly higher for commercial than residential buildings since commercial buildings start at a lower baseline electrification. (b) The flexible load is the percentage of total energy that can be reallocated throughout the day. (c) The DER energy penetrations are percentages of the average daily energy demand, including from EV charging. (d) Relative magnitudes of various energy components for 3 networks in 2050.}
\label{fig:p_years}
\end{figure*}

The aggregated projections over all networks of DER energy penetrations and electrification which we use in our simulations are given in Figure~\ref{fig:p_years}. Using these projections and the procedures detailed in the Experimental procedures section, we first compare the impacts of projected increases in DER penetrations and electrification on grid reliability with and without coordination and quantify the increase in consumer electricity costs due to DER coordination. Next, we show the potential reduction in peak network load achievable with coordination. Finally, we explore the impacts of the amount of: (i) flexible load available, (ii) the degree of stationary storage spread across the network, and (iii) the EV charger rated power on the results.



\begin{figure}[htpb]
\centering
\includegraphics[scale=0.59]{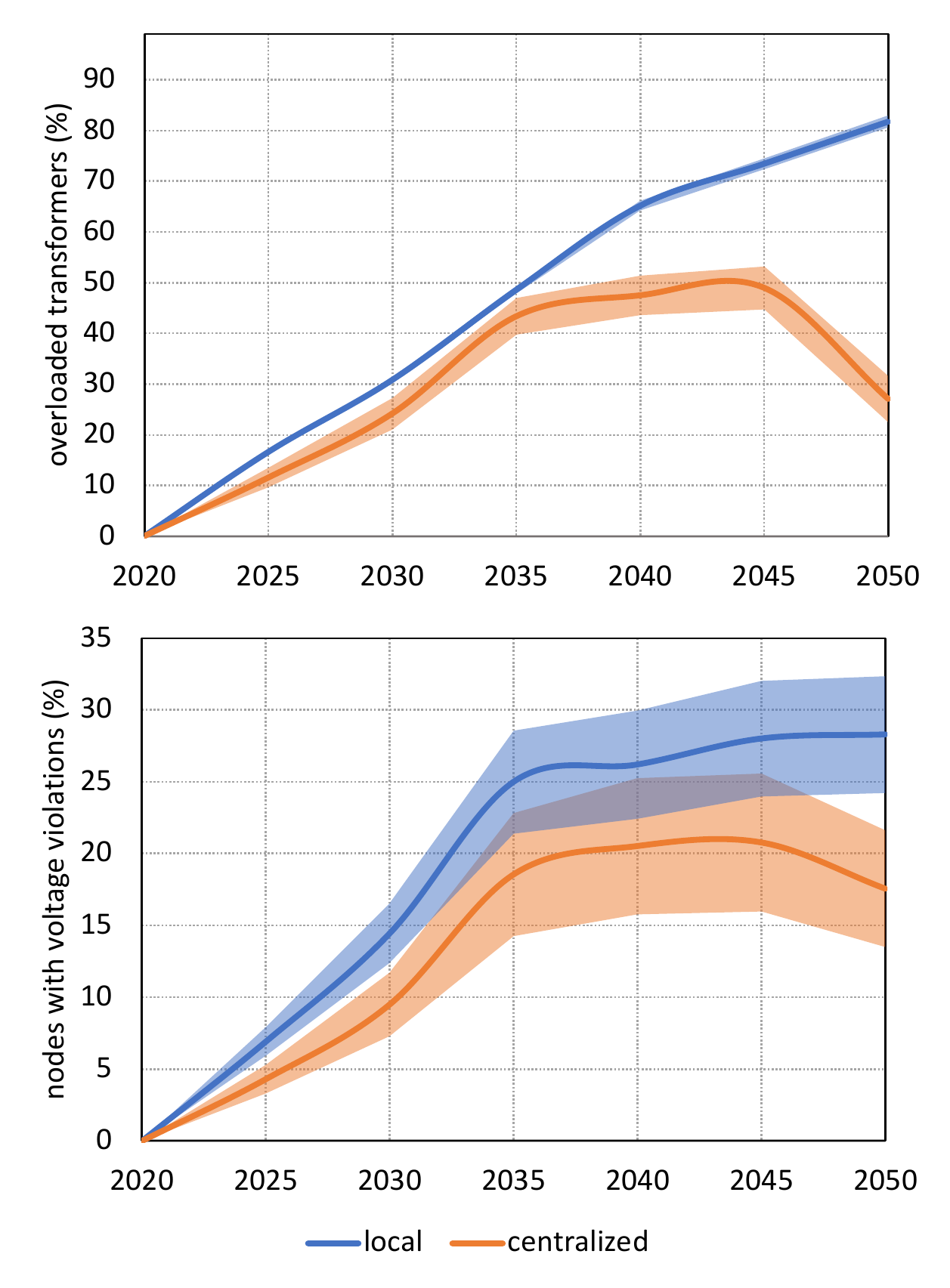}
\caption{\textbf{Computed means and standard errors of the violation percentages with both control schemes for each simulation year averaged over all networks.} Transformer overloading with centralized control grows more slowly after 2035 and drops sharply after 2045. The number of nodes with voltage violations increases rapidly until 2035 when the leaf nodes in the network are saturated with voltage violations. After 2035, the penetration of DERs increases to the point where the centralized controller can start to reduce violations significantly. After 2045 a significant portion of uncontrollable load is replaced by flexible load.
}
\label{fig:years}
\end{figure}

The aggregated projections of the violation percentages with the local and centralized controllers are given in Figure~\ref{fig:years}. As seen, in the year 2020 with low DER penetrations, there are no violations. By the year 2050 with much higher DER penetrations, electrification, and load flexibility, with local DER control, on average 81\% of the transformers experience overloading and 28\% of the nodes experience voltage violations. With DER coordination these percentages can be potentially reduced to 18\% and 29\%, respectively, which promise significant reductions in needed future grid infrastructure upgrades. Hence, with the projected increases in flexible load in 2050, there is a potential to completely eliminate 52\% of overloaded transformers and 10\% of nodes with voltage violations. Note, however, that with the centralized controller, the violations first increase until the year 2035, flatten, then drop significantly by 2050. This surprising behavior is due to the increased penetration of flexible thermal loads and storage in later years. Hence, accelerating the adoption of these DERs, for example, to the level of 2050 would help realize the full potential of DER coordination to reduce infrastructure upgrades. 
The main reason for the voltage violations being much lower than the transformer violations is the nodes close to the substation do not suffer from voltage violations. 


\begin{figure}[htpb]
\begin{tabular}{cc}
\begin{subfigure}[b]{0.5\textwidth}
\includegraphics[scale=0.45]{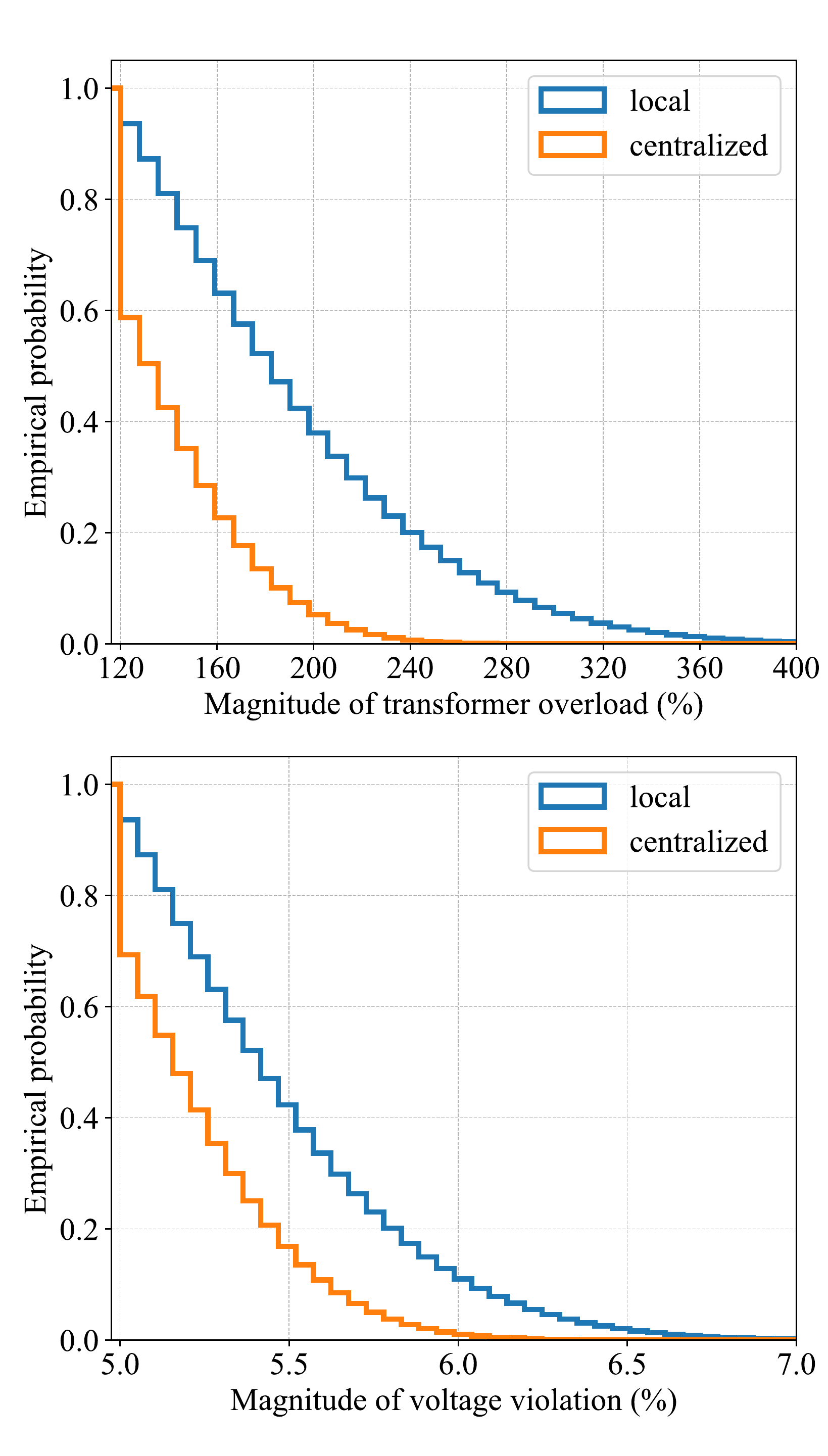}
\end{subfigure}
\end{tabular}
\caption{\textbf{The empirical probability of the percentage magnitude of transformer apparent powers or node voltages being greater than x across all networks and scenarios in 2050.} In addition to completely removing many violations with centralized control, the average magnitude of remaining transformer and voltage violations drops from 243\% to 159\% and from 5.93\% to 5.32\%, respectively. Note that shifting the distribution of the magnitudes of all transformer apparent powers and voltage towards the y-axis represents less violations.
}
\label{fig:hist}
\end{figure}

Since our simulations provide the apparent power through every transformer and the voltage magnitude for every node, we can obtain a more refined view of how DER coordination improves reliability beyond the aforementioned results using the binary-valued reliability metrics. Figure~\ref{fig:hist} plots histograms of the percentage magnitudes of the deviations of transformer apparent power and node voltage from their respective nominal values for all networks and scenarios for the year 2050. In addition to eliminating violations for significant fractions of transformers and nodes, centralized control reduces the magnitude of the remaining violations, especially the largest ones, which could help further reduce required grid infrastructure upgrade costs.

\begin{figure*}[htpb]
\centering
\begin{tabular}{cc}
\begin{subfigure}[b]{0.5\textwidth}
\centering
\includegraphics[scale=0.66]{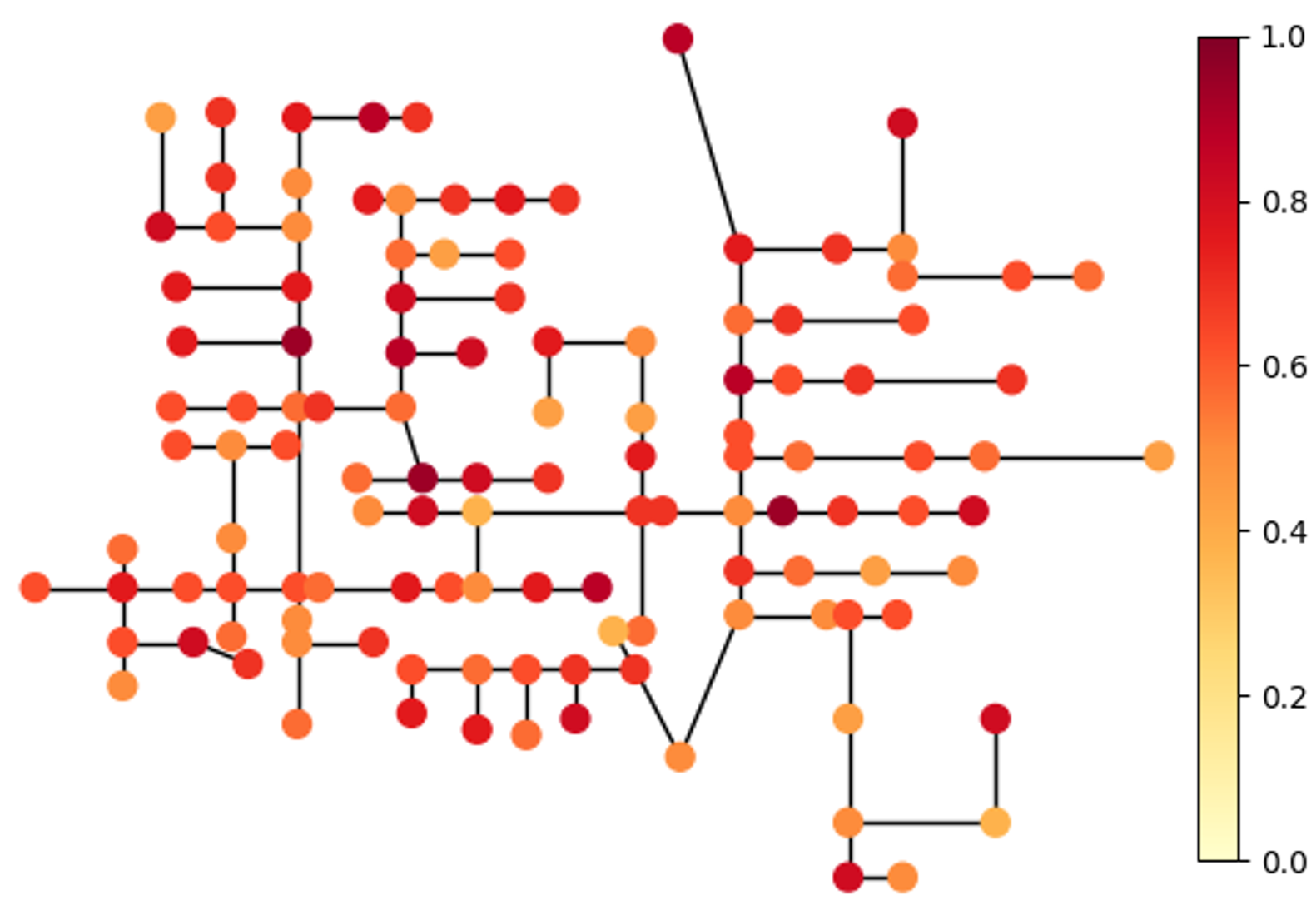}
\caption{Overloaded transformers with local control in 2040.}
	\label{fig:topology_t1}
\end{subfigure}
&
\begin{subfigure}[b]{0.5\textwidth}
\centering
\includegraphics[scale=0.66]{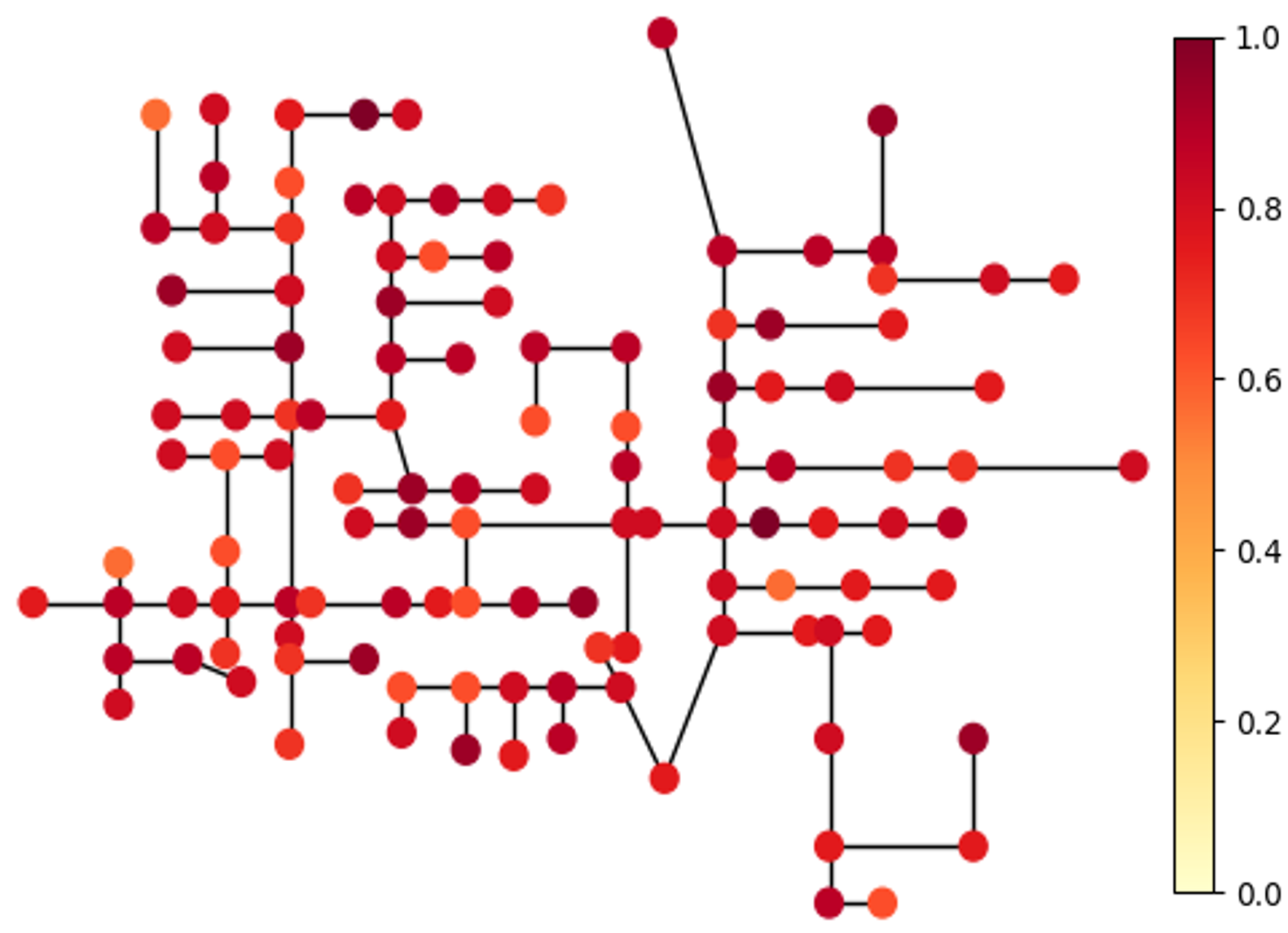}
\caption{Overloaded transformers with local control in 2050.}
\label{fig:topology_t2}
\end{subfigure}
\\
\begin{subfigure}[b]{0.5\textwidth}
	\centering
	\includegraphics[scale=0.66]{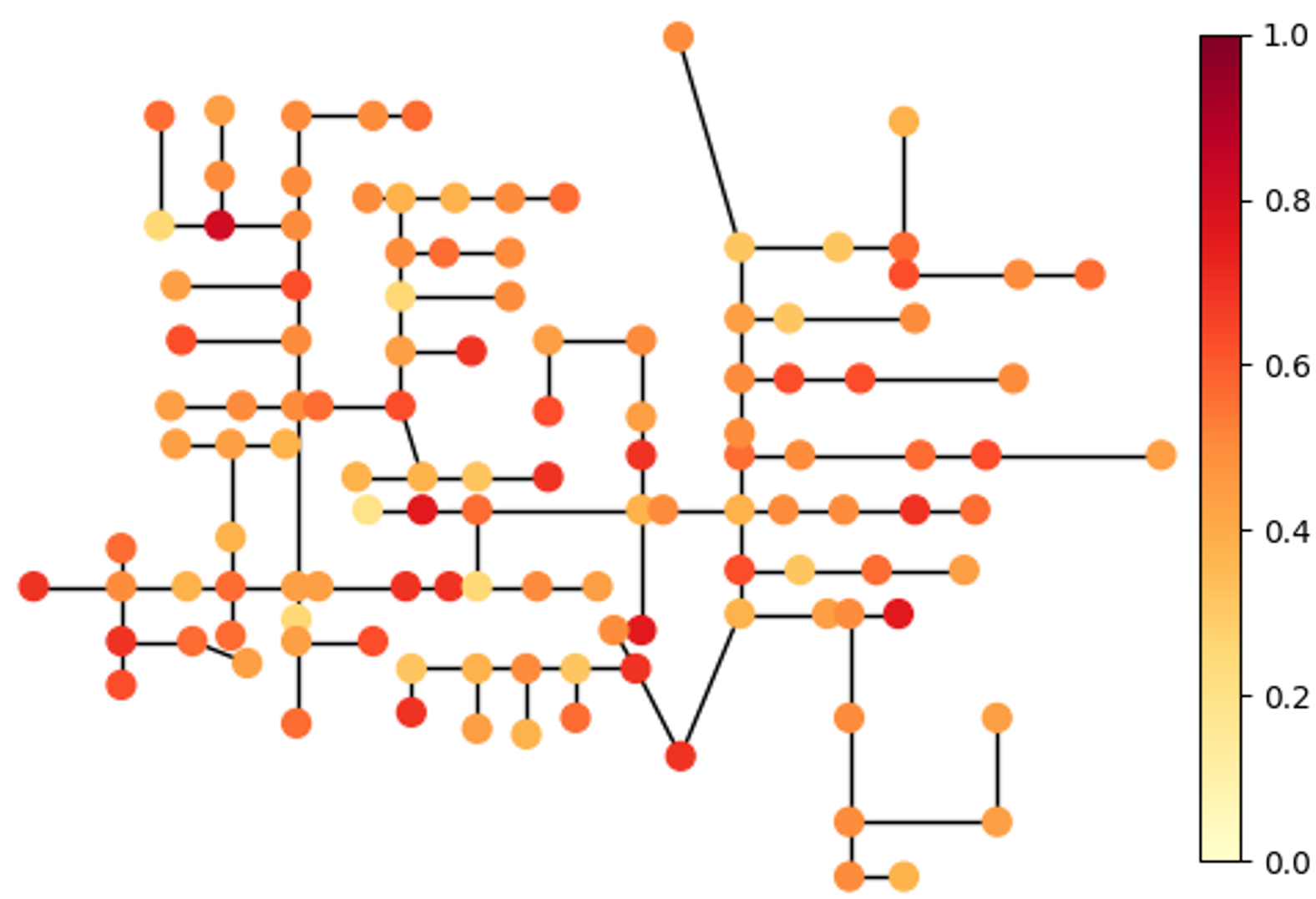}
    \caption{Overloaded transformers with centralized control in 2040.}
	\label{fig:topology_t3}
\end{subfigure}
&
\begin{subfigure}[b]{0.5\textwidth}
	\centering
	\includegraphics[scale=0.66]{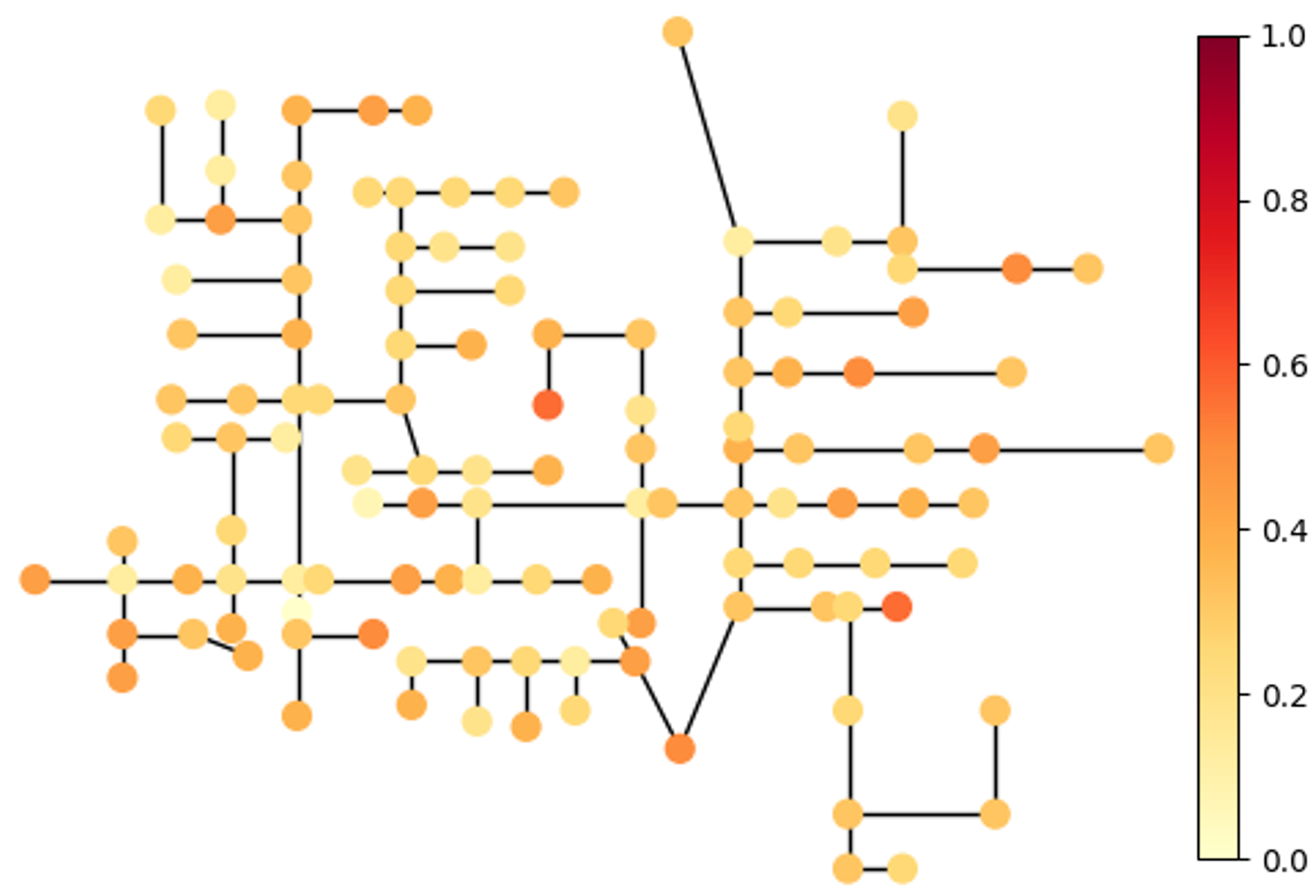}
    \caption{Overloaded transformers with centralized control in 2050.}
	\label{fig:topology_t4}
\end{subfigure}
\end{tabular}
\caption{\textbf{Topological heatmaps of the fraction of scenarios each transformer is overloaded with the local and the centralized controllers for the Sacramento network in 2040 and 2050.} Each node in the network represents an aggregation of consumers behind a secondary transformer. The transformers with the most frequent overloads have higher concentrations of EV chargers. With the local controller, in 2040 and 2050 62\% and 75\% of the transformers experience overloading in half or more of the simulation scenarios, respectively. With the centralized controller, these numbers are reduced to 47\% and 26\%.}
\label{fig:topologies}
\end{figure*}

Zooming into the simulation results for a network, Figure~\ref{fig:topologies} depicts topological heatmaps of transformer violations with the local and centralized controllers for the Sacramento network in 2040 and 2050. With the local controller, transformer violations increase from 2040 to 2050. However, with the centralized controller, transformer violations in this network decrease from the peak in 2040 to much less in 2050 due to the increase in flexible load and storage. As seen, some of the transformers that had a high probability of being overloaded in 2040 have their probabilities significantly reduced in 2050. The information provided by such topological heatmaps can be used, for example, to plan specific future grid infrastructure upgrades.

\smallskip

\subsection*{Impact on electricity cost} 
Since the centralized controller prioritizes grid reliability over minimizing electricity cost, it is expected to yield higher costs than with the local controller whose main objective is to reduce consumer electricity costs. Using the 2050 cost results for the 11 networks, we found that the centralized controller increases the consumer electricity cost between 4.7\% and 5.6\% over the local controller. The reason for the modest cost increase with centralized control is that all the flexibility for the customer in the form of storage, EV charging, and thermal load is used for cost reduction while the centralized controller uses a small portion of it for grid reliability. 

Note that this cost increase includes only the energy cost portion of the consumer electricity bill according to current TOU retail rates. A significant portion of the total electricity cost to the consumer also includes the maintenance costs, capital investment, and asset depreciation of the distribution grid. There is previous work highlighting the substantial magnitude of the potential future distribution grid costs to accommodate increased electrification~\cite{BCG, EIA_grid_cost}. However, due to the complexity of predicting future energy and distribution asset costs and their wide variations across different regions, it is a question of further research exactly how the savings of distribution maintenance and asset costs will compare to the potential increase in energy generation costs in the future.



\subsection*{Impact on peak network load}

\begin{figure}[htpb]
\hspace{-15pt}
\includegraphics[scale=0.65]{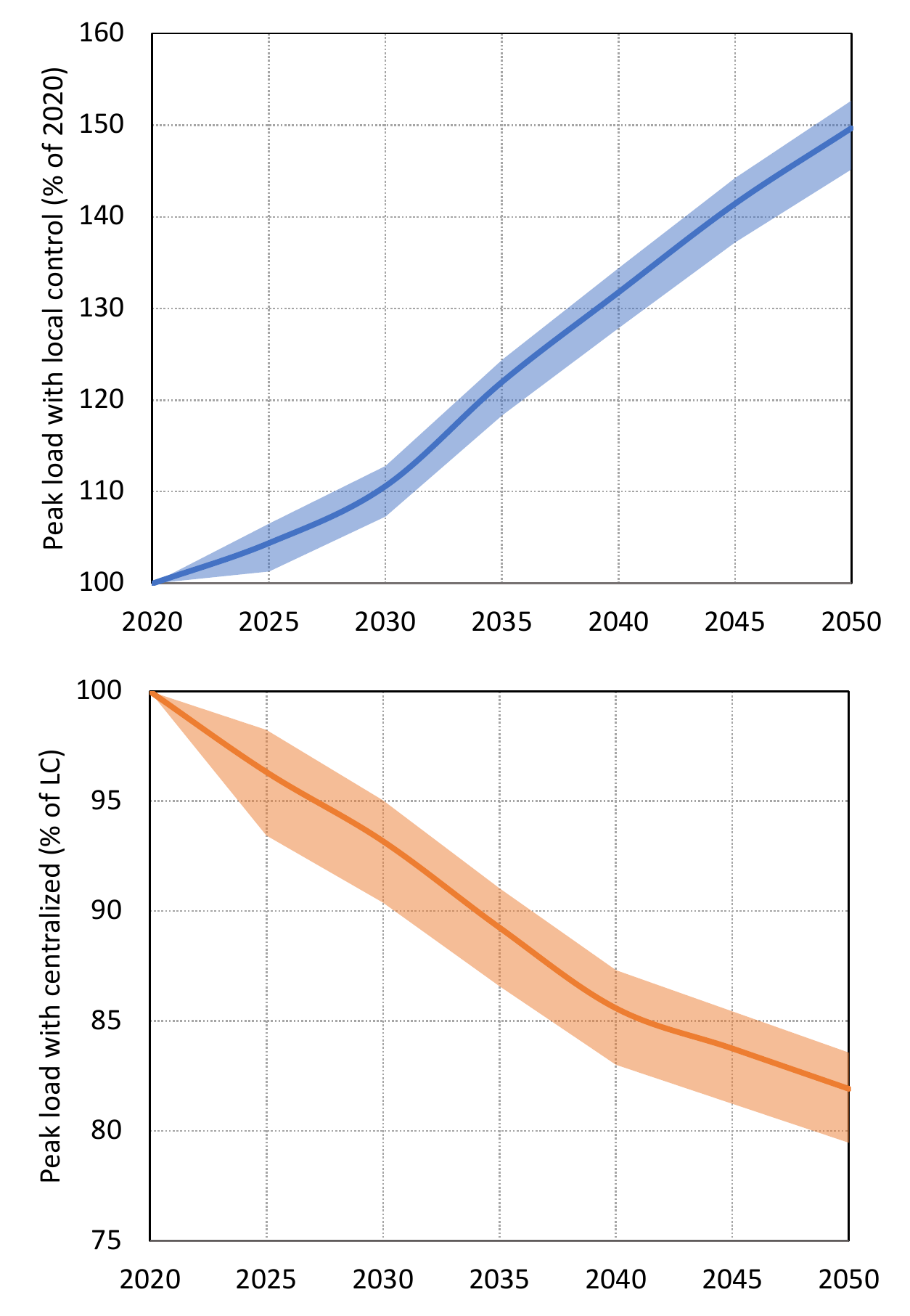}
\caption{\textbf{Plot of the means and standard errors of the peak load over the simulation horizon with local control relative to 2020 and centralized control relative to local control.} Top: Means and standard errors of peak load with local control averaged over all networks from 2020-2050. As the EV charging and thermal load electrification increase, the peak load increases with local control. Bottom: Means and standard errors of peak load with local control averaged over all networks from 2020-2050. As DERs and load flexibility increases, the centralized controller reduces peak load further relative to the local controller.
}
\label{fig:peak_load}
\end{figure}

The primary goal of current DER coordination programs, such as VPP and demand response, is to reduce total network peak load with no consideration of grid reliability. In contrast, the goal of the centralized controller in this study is to reduce reliability violations with no explicit attempt to reduce network peak load. Does it also reduce peak load? To answer this question, we compare peak load over the simulation horizon with our centralized controller to that with the local controller for all networks. As can be seen in Figure~\ref{fig:peak_load}, with the local controller, peak load is projected to increase on average by 22\% in 2035 and 50\% in 2050 over the 2020 baseline due to the increased adoption of DERs and electrification. With the centralized controller, average peak load is reduced by 12\% in 2035 and 18\% by 2050. Hence, coordinating DERs for grid reliability can also reduce peak load. This does not mean, however, that reliability coordination should completely replace demand response and VPPs, since such programs would still be needed to deal with extreme climate or other conditions in which peak load must be more significantly reduced.

\subsection*{Impact of flexible load penetration}

\begin{figure}[htpb]
\centering
\hspace{-15pt}
\includegraphics[scale=0.65]{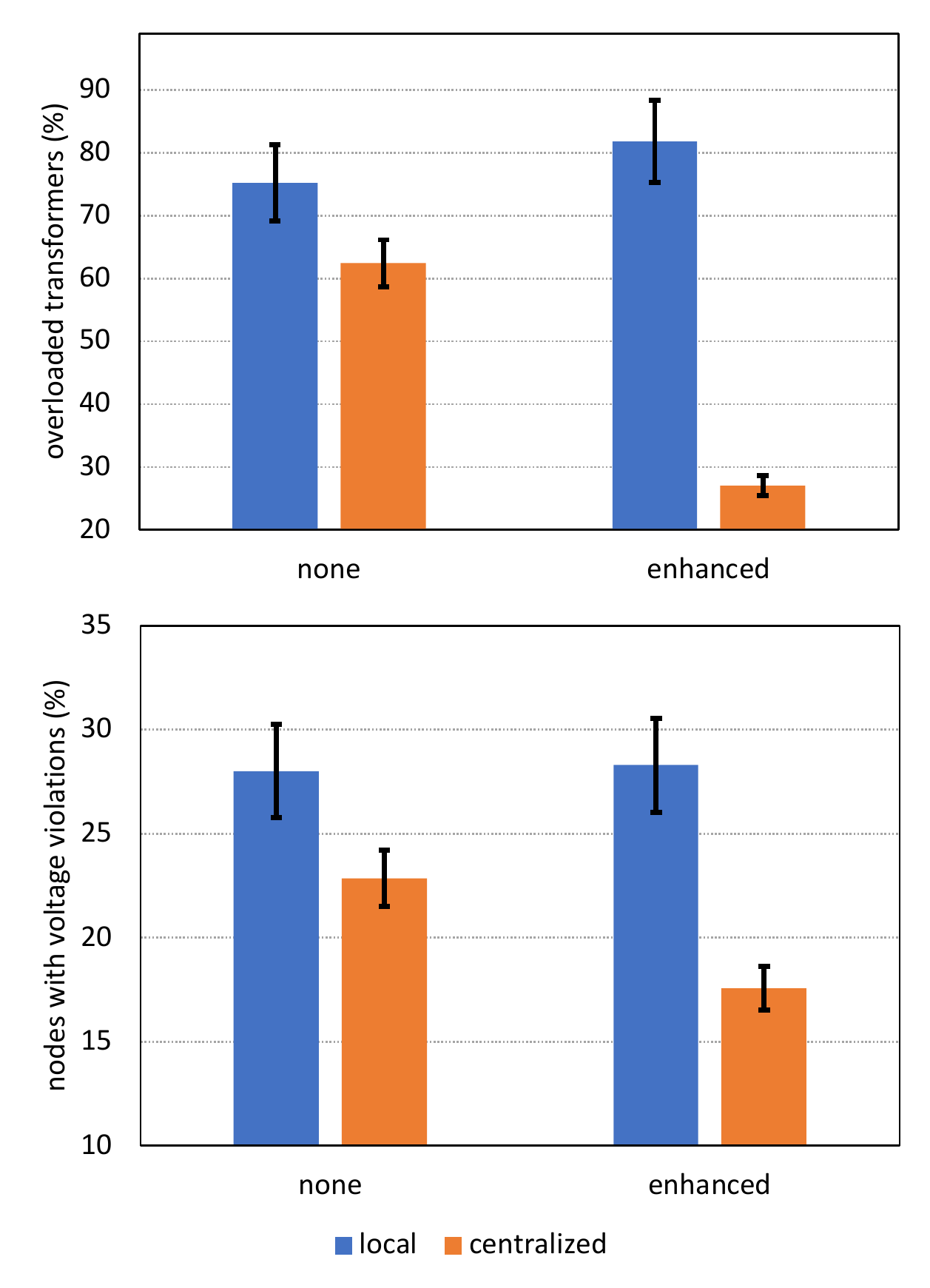}
\vspace{-15pt}
\caption{\textbf{Snapshots for the year 2050 of the percentages of transformers with overloading and nodes with voltage violations with local DER control and perfect foresight centralized control with none and enhanced flexible load control.} With centralized control, the percentages of transformers with overloading and nodes with voltage violations are reduced on average from 80\% to 29\% and from 28\% to 18\%, respectively, depending on the network. With local control, the number of nodes with overloaded transformers and voltage violations is 2.83 and 1.58 times higher than those with centralized control, respectively. This implies the cost of upgrades with or without coordination would have a similar ratio. Without flexible load control, the transformer violations with the local controller are less by 7\% on average, indicating that the local controller mismanages the flexible load when it comes to grid reliability. However, the centralized controller can only reduce transformers with overloading and the nodes with voltage violations by 15\% and 4\% respectively.}
\label{fig:bars_flex}
\end{figure}

The projections of flexible load in~\cite{flex_load_NREL} include a base and enhanced case. While we assumed the enhanced case in the reported results in this study, we also simulated the impact with no available load flexibility. Figure~\ref{fig:bars_flex} shows that the centralized controller is able to further reduce transformer and voltage violations when there is enhanced load flexibility.
Without thermal load flexibility, there are high percentages of transformer and node voltage violations both with local control and with centralized control. 
With full thermal load flexibility, violations actually increase slightly under local control but improves cost by 1.9\%. This increase in violations with local control occurs because the flexible load and EV charging are shifted just outside the peak price period resulting in large power consumption spikes. Without load flexibility, the thermal load shape is more spread out across the peak and non-peak time periods. On the other hand, centralized control dramatically reduces violations by 34\% and 6\% for transformers and voltages respectively on average while improving cost by 1.1\%.
Hence, centralized control is essential in realizing the full benefits of thermal load flexibility.

\subsection*{Impact of stationary storage spread}
Stationary storage can help improve network reliability by shifting load away from peak periods and storing excess solar PV generation. The projections we used from~\cite{NREL_storage}, however, assume that storage is spread over only a fraction of the nodes with solar, e.g., 70\% in 2050. What is the impact of varying this spread on grid reliability? To answer this question, we also compare the violations when the same projected total amount of storage in the network is spread over 100\% of the nodes with solar. 

\begin{figure}[htpb]
\centering


\hspace{-15pt}
\includegraphics[scale=0.65]{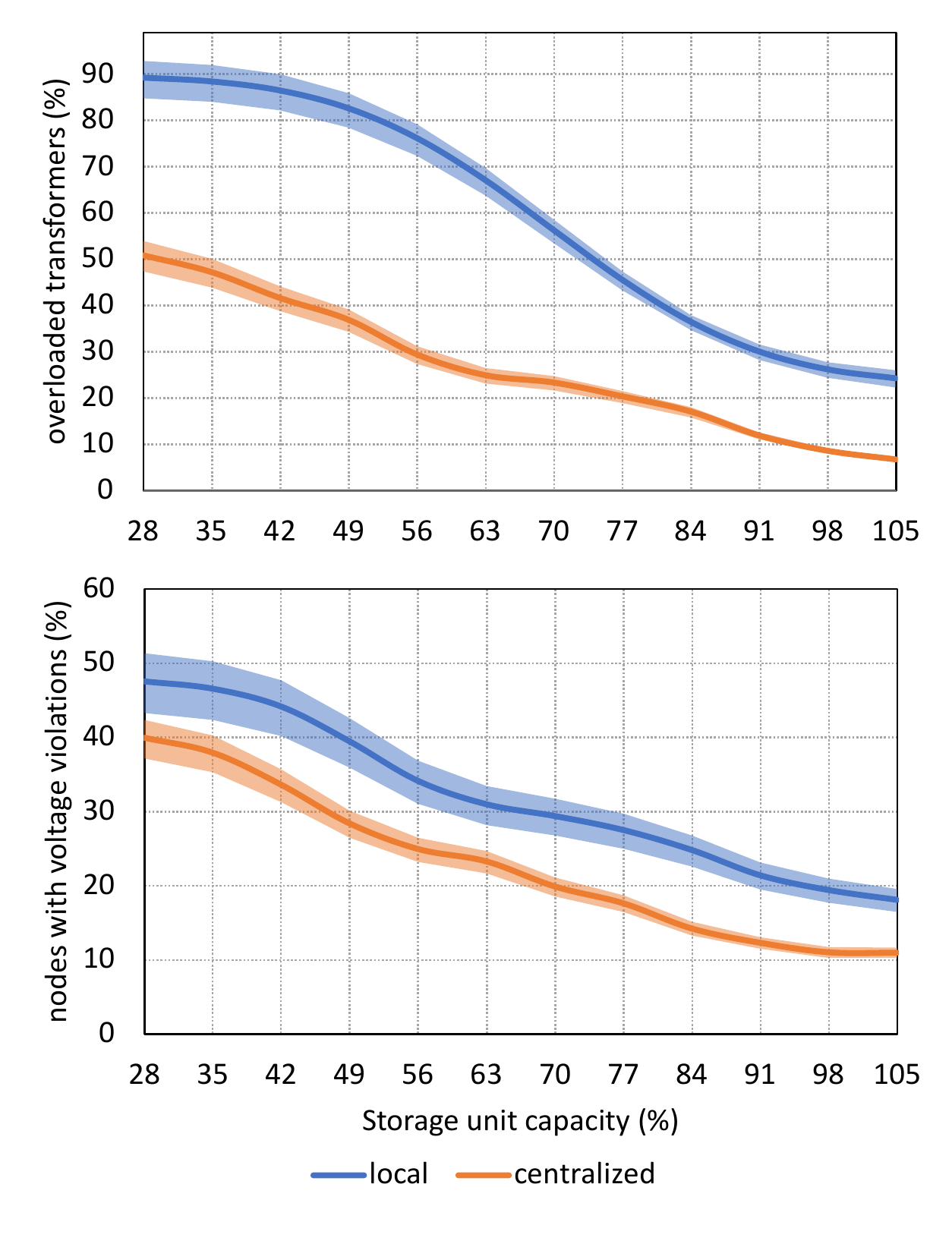}\\
\caption{\textbf{Plots of the violation percentages under varying amounts of total network storage capacity for 100\% storage spread in 2050 for the Tracy network with the local and centralized controllers.} The storage capacity is a percentage of the projected average consumer storage capacity in 2050, which is 10.1kWh. The storage capacity with the maximum rate of decreasing violations for the centralized controller occurs between 56-63\% and 49-56\% for transformers and node voltages, respectively.}
\label{fig:stor_capacity}
\end{figure}

One way to view the advantage of a wider spread of storage is in terms of the potential reduction in total storage capacity for a given violation percentage. To exemplify such reduction, Figure~\ref{fig:stor_capacity} plots the violation percentages for the Tracy network at 100\% storage spread for several average consumer storage capacities expressed as percentages of the average projected consumer storage capacity. From Figure~\ref{fig:bars_flex}, under the projected consumer storage capacity of 8.3kWh, the amount of overloaded transformers are 82\% and 28\% for the local and centralized controllers, respectively. For these same transformer overload percentages, Figure~\ref{fig:stor_capacity} gives an average consumer storage of around 5.39kWh and 5.54kWh.

\subsection*{Impact of EV charger rated power} 
\begin{figure}[h]

\centering
\includegraphics[scale=0.45]{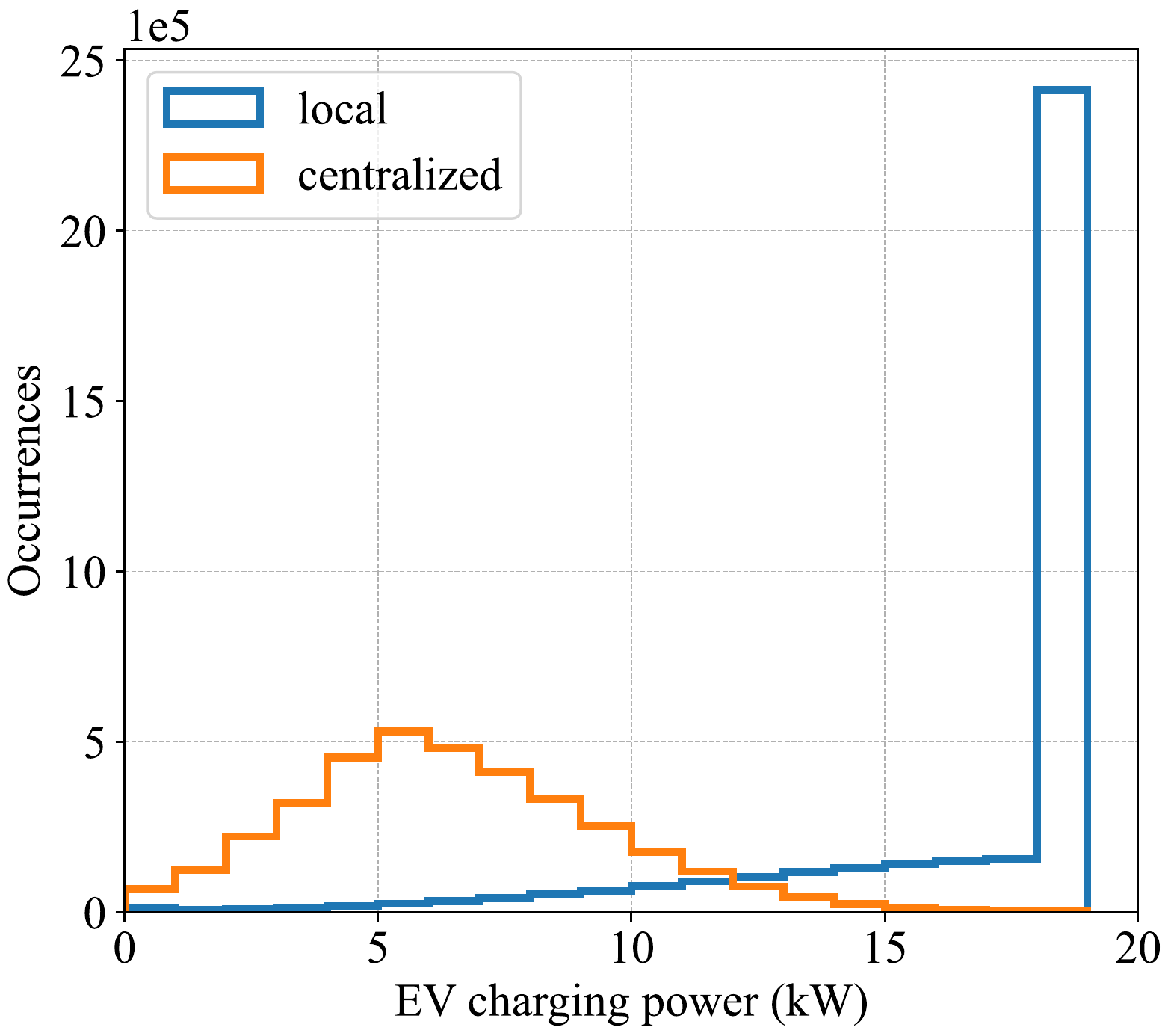}
\caption{\textbf{Histogram of EV charging power across all simulations in the year 2050 with 19kW maximum charging power.} The local controller defaults to charging at the maximum power during low price periods and only charges at lower power when not as much EV charging is needed. In comparison, the centralized controller limits charging power to avoid grid violations as much as possible while still fully charging the EV in the allowed window.}
\label{fig:evs}
\label{fig:ev_hist}
\end{figure}

Our previous results assumed L2 EV charger rated power of 6.3kW. To speed up EV charging, higher rated power chargers of 12kW and 19kW are being proposed~\cite{EV_charging_CEC}. What are the impacts on these higher power chargers on our results? To answer this question we repeat our experiments for each of these higher powers. The difference between the performance of the two controllers at higher power levels is explained in Figure~\ref{fig:ev_hist}, which plots histograms of EV charging power for all networks and scenarios with EV charger rated power of 19kW. Note that the average charging power for the local controller is 18.5kW compared to only 6.8kW for the centralized controller. This is due to tendency of the centralized controller to spread charging over higher fractions of the available windows, which is often many hours due to residential owners charging overnight. Hence, with coordination there is little benefit to widespread adoption of 19kW charging as there are only a few opportunities to make use of such high power without harming grid reliability.

Changing EV charging rated power to 19kW increases the cost of the centralized controller further to 5.5-6.1\% over the local controller because to minimize cost, the local controller leverages the higher charger power to charge more during the low price periods.
Given that we only consider the TOU energy cost of the electricity bill, this increase should not be a major barrier to adoption of DER coordination. 

\section*{Key findings}


Using a suite of distribution networks and recent projected increases in DER and electrification penetrations, a power flow driven simulation and optimization methodology is used to quantify the impacts of these increases on distribution grid reliability over the next three decades. Our main findings are:
\smallskip

\noindent{1.} Using current local DER control, on average 81\% of the transformers and 28\% of the nodes are expected to experience violations by 2050, requiring major and costly grid infrastructure upgrades. Using the perfect foresight centralized DER controller, these percentages are reduced to 28\% and 18\%, respectively, in 2050, which promise major savings in required infrastructure upgrades. We found that with the centralized controller, the violations first increase then decrease with increased adoption of flexible load and storage. These violation reductions are achieved with an average increase in consumer electricity cost of only 5\% over the local controller whose objective is to reduce consumer electricity cost. 

\noindent{2.} Although the centralized controller's main objective is to reduce violations, it significantly reduces total grid peak load\,--\,an important goal in overall grid decarbonization\,--\,by 7\% to 24\% in 2050. This suggests that coordinating DERs for grid reliability would complement the current programs, such as demand response and VPP programs, which aim to reduce electricity costs during high grid peak load events but with no consideration of grid reliability.

\noindent{3.} The high penetration of flexible load assumed in our study is a major contributor to the significant reductions in violations with the centralized controller. However, if reliability is not considered by the DER controller, as is the case today, thermal load flexibility actually increases violations. Therefore, future schemes to integrate and incentivize thermal load flexibility should emphasize reliability as an objective.

\noindent{4.} The centralized control can help reduce violations even more over local control for networks with higher spread of storage across the nodes with solar PV and networks with higher EV charger rated output power.

The results of our study suggest that future programs for incentivizing the adoption and integration of DERs, including thermal loads, EVs, solar PV and stationary storage, should include a requirement for consumers to participate in reliability coordination in order to avoid the violations seen with lower adoption of flexibility. Since addressing grid reliability requires some detailed information about the grid, the distribution system operator (DSO) will need to play a key role in such coordination programs.
\section*{Discussion}

While our results suggest that DER coordination can provide significant benefits to grid reliability, it is not clear how much of these benefits can be attained by practical coordination schemes, which must operate under temporal and spatial uncertainties of load and renewable generation, incomplete information about the network, communication network constraints, and more importantly the fact that DERs are owned by different consumers with varying objectives and privacy constraints. The development of implementable schemes for DER coordination under these constraints remains a significant technical challenge~\cite{kyle_anderson:2017,Navidi_smartgridcomm}. 

As in any data-driven study, there are several limitations stemming mainly from lack of sufficient data and justifiable assumptions. Limitations of our study include: (i) EV charging events represent only current SF Bay area patterns, (ii) there is always a sufficient number of EV chargers at each commercial consumer node to accommodate charging demand without any queuing, and (iii) except for dependence on node load and type, the distribution of DERs across the nodes is assumed to be uniform which may not be realistic given possible variations in consumer demographics. Additionally, since we simulate the average trend of DER penetrations, it is possible that outlier networks will have radically different penetrations of DERs, so the conclusions would not apply to them. When considering how other combinations of DER penetrations may impact the results, one must consider how some penetrations increase the flexibility while others increase the load. As the load increases, DER flexibility becomes more important to ensure reliability. On the other hand, if the flexibility is dramatically reduced, the benefit of coordination would be less as you are closer to the 2020 penetration scenario.

Our analysis can be extended in several directions. One extension would be to include other reliability metrics that are amenable to quasi-static power flow analysis, such as constraints for power lines, breakers, and voltage regulators. Another extension would be to compare the results under different electricity tariffs. A potentially more important extension would be including net-metering and DER participation in ISO markets, such as wholesale or ancillary service markets, as suggested by the FERC order 2222. However, our results do not change when considering a net-metering scheme that provides a lower compensation than off-peak TOU rate, for example, wholesale market rate. Some current net metering schemes compensate net-metering at the TOU peak price, effectively providing sufficient free storage to offset all excess solar. In this case, consumers have no incentive to deploy stationary storage, which harms grid reliability through excessive solar backfeeding. While our methodology can accommodate DER participation in ISO markets, such participation is highly dependent on the portfolio of transmission system assets and the prices of such services. This makes the design of a scenario and control scheme to use in our methodology too speculative at this time. 

Finally, the framework presented in this paper can be used to benchmark the impact of new technology options and investments under consideration in ongoing electrification programs, such as, increased adoption of heat pumps, water heaters and vehicle to grid charging.


\section*{Experimental procedures}

\subsection*{Resource availability}
\subsubsection*{Lead contact}
Further information and requests for resources should be directed to the lead contact, Thomas Navidi (tnavidi@stanford.edu)

\subsubsection*{Materials availability}
This study did not generate new unique materials

\subsubsection*{Data and code availability}
All data and data generators used in the study are publicly available and referenced in this paper. The code used to run the experiments in this paper is available at: \url{https://github.com/tnavidi1/DGSS}

\subsection*{Simulation methodology}

The organization of this section generally follows the block diagram in Figure~\ref{fig:block_diagram}. The numbers and letters in the block diagram are referred to when applicable. Pseudo-codes of some of the more complex procedures and further details are provided in the Supplement.

\subsubsection*{Network model, load profiles}
\noindent{(Figure~\ref{fig:block_diagram} A)}. The 3-phase distribution network model of each network provides the information needed to perform power flow simulation using OpenDSS~\cite{montenegro2012real}, including topology, line impedance values, windings and electrical characteristics of transformers. The network model also includes average daily peak real power consumption for every consumer node. Some networks have their transformer rated power capacities specified. For the other networks, we set the capacities to 120\% of their respective peak values in the 2018 dataset.
As in OpenDSS, a node in our work refers to a bus-phase combination and each of the three phases having separate power lines. Hence, a consumer site can be modeled by more than a single consumer node, each having its own independent load. Most of the consumer sites in our study have a single phase or are balanced over multiple phases. Similarly, a 3-phase transformer is modeled by 3 single phase transformers each with its own apparent power capacity.
Table~\ref{tab:network_1} shows the main characteristics of each of the distribution grids (see Supplement for additional characteristics).

\begin{table*}[!htb]
\caption{\textbf{Main characteristics of the networks used in the simulations.} The table shows the assigned name indicating location and source of network data, population density type, the numbers of voltage nodes $\boldsymbol{N}$, number of consumer nodes or load points $\boldsymbol{N}_\mathrm{C}$, \% of consumer nodes that are commercial $\boldsymbol{n}_\mathrm{CC}$, number of transformers $\boldsymbol{N}_\mathrm{T}$, and average daily peak power $\boldsymbol{\hat P}$ in MW for each network. The asterisk by the network name indicates a network that aggregates consumers under the secondary transformer. The other networks have synthetic models for each individual consumer below its secondary transformer.}
\label{tab:network_1}
\vspace{3pt}
\centering
\begin{tabular}{llllllllllll}
\hline\\[-9pt]
\textbf{Name}                                                    & \textbf{Type} && $\boldsymbol{N}$&& $\boldsymbol{N}_\mathrm{C}$ && $\boldsymbol{n}_\mathrm{CC}$ && $\boldsymbol{N}_\mathrm{T}$ && $\boldsymbol{\hat P}$ \\[2pt] \hline\\[-8pt]
Sacramento*~\cite{test_feeders_IEEE}     & Synthetic, Suburban      && 278                  && 91                            && 9                                   && 99                          && 4                           \\[2pt]
Iowa*~\cite{iowa}                          & Real, Suburban      && 915                  && 193                           && 6                                   && 268                         && 2                           \\[2pt]
Central SF~\cite{grid_models_NREL}       & Synthetic, Urban         && 2115                  && 425                           && 21                                  && 232                         && 8                           \\[2pt]
Commercial SF~\cite{grid_models_NREL}    & Synthetic, Urban         && 172                   && 18                            && 100                                 && 17                          && 2                           \\[2pt]
Tracy~\cite{grid_models_NREL}            & Synthetic, Suburban      && 775                  && 161                           && 13                                  && 108                          && 2                           \\[2pt]
Rural San Benito~\cite{grid_models_NREL} & Synthetic, Rural         && 243                   && 22                            && 10                                  && 15                           && 0.1                            \\[2pt]
Los Banos~\cite{grid_models_NREL}        & Synthetic, Suburban      && 2010                  && 426                           && 11                                  && 251                         && 2                           \\[2pt]
Vermont~\cite{EPRI_J1}                    & Real, Suburban      && 4245                  && 1384                           && 9                                   && 828                         && 6                           \\[2pt]
Arizona*~\cite{test_feeders_IEEE}        & Real, Suburban      && 138                   && 38                            && 14                                  && 46                          && 2                           \\[2pt]
Marin~\cite{grid_models_NREL}            & Synthetic, Suburban      && 3689                  && 811                           && 10                                  && 231                         && 3                           \\[2pt]
Oakland~\cite{grid_models_NREL}          & Synthetic, Suburban      && 10073                 && 2426                          && 13                                  && 658                         && 12                           \\[2pt] \hline
\end{tabular}
\end{table*}

\noindent{(Figure~\ref{fig:block_diagram} B)}. To construct the baseline load profiles, we use residential and commercial customer data from~\cite{NREL_stock_data} which is modeled after measured data collected in 2018. For each network, we extract hundreds of 1-year, 15-minute load profiles from this dataset based on its geographical census code. Although some networks provide load profiles, we do not use them to account for the electrification of thermal loads. Instead, we use the average daily peak load from the provided data to calibrate load at each node to ensure compatibility. The rest of the load profile data comes from the NREL dataset, so that we can capture the different components of electrification. To set the baseline load profile for each consumer node in the network, we randomly select one load profile– from the extracted data which has average daily peak load within $\pm 10\%$ of that for the node. The randomly selected profile determines whether the node corresponds to a residential or commercial consumer. This baseline load profile is then scaled so that its average daily peak load matches the value given by the network.

Projections of the increased energy due to electrification in the US up to 2050 are given in~\cite{NREL_EV_adoption}. We consider the high scenario with estimates of residential and commercial for 2050 of up to 14\% and 39\%, respectively, relative to their corresponding 2018 demand, depending on the state and categorized by space heating, water heating, and other. To determine the load profiles for a future network scenario, we use the following procedure: First, the electrification due to loads other than thermal are uniformly increased by their electrification projection percentage. We then randomly choose a consumer that does not already have electric space heating or water heating, and assign to it a load profile of either space or water heating. For each selected consumer, we convert its thermal energy profile from~\cite{NREL_stock_data} to electric energy using the median published efficiency for the corresponding electric appliance in~\cite{CEC_appliance}. The electric energy profile is then added to the existing baseline load profile for the consumer. This process is repeated until the total added electric energy in a network equals its electrification projection. The reactive power component of the load for each node is chosen such that the power factor is randomly distributed between 0.9 and 0.95 as is commonly assumed, e.g., see~\cite{iowa}.


\subsubsection*{DER parameters and penetrations}
\noindent{(Figure~\ref{fig:block_diagram} C)}.
Workplace and home L2 chargers are assumed to have a maximum charging power of 6.3kW~\cite{NREL_EV}. We assume that the charging power can be set to any value between 0 and the maximum. We do not include EV fast charging stations for two reasons. First, fast charging does not have flexibility, thus a limited amount of fast charging would be similar to how we model the other inflexible loads that increase the peak load at the added node. If a lot of fast charging is added to one node, the utility would likely add a corresponding upgrade to that grid or use storage to offset the load. Any new infrastructure added to support a DC fast charger would unlikely experience violations as it would be built to handle them. Additionally, the use of added storage to offset the fast charger would be similar to how we already model storage and the other inflexible loads.

Each storage unit has a maximum c-rate of 0.5~\cite{NREL_storage} and a round-trip efficiency of 86\%~\cite{storage_efficiency}. 

\noindent{(Figure~\ref{fig:block_diagram} D)}. EV penetration is defined as the EV charging energy as a percentage of the total energy in the network with no solar or EVs included. The penetrations we assume are derived from the projections in~\cite{NREL_EV_adoption}. 
We consider the high electrification due to EVs scenario in 2050 of 29\% to 38\% relative to the 2018 baseline load depending on the state.


PV penetration is the energy generated by solar PV as a percentage of total energy consumed by the network, including for EV charging. We set the PV penetration in 2050 for each network to 23\% of the total potential rooftop solar PV generation of its nearest city~\cite{NREL_solar},
which corresponds to approximately the same amount of solar PV capacity as the high penetration scenario for nationwide distributed PV projected in~\cite{NREL_storage}. To determine the PV penetration for the years prior to 2050, we scale the penetrations in proportion to the nationwide PV capacity provided in~\cite{NREL_storage}. The PV generation profiles for each network are obtained from the solar data for the closest geographic region to the network provided in~\cite{NREL_solar_data}.

We assume that only nodes with solar PV can have stationary storage and define stationary storage adoption penetration as a percentage of the nodes with PV. The storage penetrations we assume correspond to the highest adoption rates given in~\cite{NREL_storage}, which range between 30\% and 70\% depending on the year. 

\noindent{(Figure~\ref{fig:block_diagram} E)}. The TOU tariffs~\cite{pge_tariffs} include peak, part-peak, and off peak rates for both residential and commercial consumers. The hours during which these rates apply are chosen such that the middle of the peak price hour corresponds to the most frequent total network peak demand hour. In addition, EV TOU rates~\cite{EV_charging_behaviors} are included; see Supplement for details.

\subsubsection*{Scenario generation}

\noindent{(Figure~\ref{fig:block_diagram} 1)}. The first step for generating a simulation scenario is to determine the EV charging windows. To do so, we use the synthetic data generator in~\cite{EV_model_siobhan} to produce sample charging schedules each consisting of a starting time, ending time, charging energy demand, and whether charging is done at home or at work over the 1-year simulation horizon. We assign the EV charging schedules to residential and commercial nodes at random but with the constraint that each residential node can accommodate at most one charger per 12.2kW of its peak load to prevent an unrealistic pileup of chargers on a few nodes. The probability of a commercial node being assigned to an EV is proportional to its peak load. The EV charging power injection profile is determined by the control algorithm used. However, the total amount of charging energy is predetermined by the data generator in~\cite{EV_model_siobhan}. We continue to add new EVs to the network in this manner until the total energy consumed by EVs is equal to the total projected EV electrification penetration energy.


\noindent{(Figure~\ref{fig:block_diagram} 2)}. Solar PV is assigned to nodes as follows. We pick a node at random and assign it a PV generation profile that is scaled such that the ratio of solar PV energy generation to the total energy consumed by the node is randomly distributed between $40-90\%$, which roughly corresponds to the upper quartile estimate of residential PV system sizes given in~\cite{NREL_storage}. 
This process is repeated until the total solar PV energy generation in the network reaches its prescribed value.

\noindent{(Figure~\ref{fig:block_diagram} 3)}. To assign storage to nodes, we first randomly select a subset of the nodes with solar PV whose size is determined by the given storage penetration. Each node in this subset is then randomly assigned a storage capacity between $40-80\%$ of the average daily PV energy generation, which corresponds to the upper quartile estimate of residential battery system sizes in~\cite{NREL_storage}. 
The battery power injection profile is determined by its control algorithm.

\noindent{(Figure~\ref{fig:block_diagram} 4)}. Finally, we assign the appropriate tariffs to each consumer node. Additionally, a randomly selected set of $40\%$ of the consumer nodes with EV chargers~\cite{EV_charging_behaviors} are prescribed an EV TOU electricity tariff for their EV charging energy; see Supplement for additional details on scenario generation.


\subsubsection*{DER control, cost computation}

\noindent{(Figure~\ref{fig:block_diagram} F, 5, 6')}. The variables and constants we use in the formulation of the control algorithms are given in Table~\ref{tab:var_tab}. We use boldface for vectors and the notation $f(t)$ to indicate a function of time step $t$. The size of a vector / matrix should be clear from the context, inequalities involving vectors are element-wise, and $\boldsymbol{p}^2$ means taking the square of each element of the vector $\boldsymbol{p}$.

\begin{table*}[htpb]
	\caption{Variables and constants used in the DER control algorithms.}
	\label{tab:var_tab}
	\begin{tabular*}{1\textwidth}{r p{11cm}}
		\hline\\[-9pt]
		\textbf{Symbol} & \textbf{Description}  \\[2pt] \hline \\[-9pt]
		$t, T, N, N_\mathrm{C}, i, j, K_i, k$  & time, number of timesteps, number of nodes in the network, number of consumer nodes, node indices, number of controllable DERs in node $i$,  controllable DER index \\[2pt]
		$\mathcal{T}, \mathcal{E}_i$ & set of input and output node pairs for all transformers, set of EVs in node $i$ \\[2pt]
		$\boldsymbol{s}, \boldsymbol{p}, \boldsymbol{q}, \boldsymbol{v}, \tau_{ij}$ & power, real power, reactive power, voltage magnitude, square of transformer apparent power magnitude \\[2pt]
		$v^{\mathrm{min}}, v^{\mathrm{max}}, {\tau}_{ij}^{\mathrm{max}}$ & limits on voltage magnitude and square of transformer apparent power magnitude \\[2pt]
		$\boldsymbol{c}, \boldsymbol{d}, \boldsymbol{Q}, \boldsymbol{Q}^{\mathrm{max}}, \boldsymbol{Q}^{\mathrm{min}}, \gamma_l, \gamma_c, \gamma_d$ & EV and storage charging power, discharging power, amount of stored energy, maximum and minimum energy capacity, battery leakage, charge efficiency, discharge efficiency \\[2pt]
        $\boldsymbol{u}^{\mathrm{base}}, \boldsymbol{u}^{\mathrm{max}}, \boldsymbol{u}, \boldsymbol{\phi}, T_d$ & base flexible load profile, maximum power consumption of flexible load, flexible load profile after control, fraction of thermal energy that can be controlled, timesteps in a day \\[2pt]
		$t^{\mathrm{end}}_k, Q_k^{\mathrm{final}}$ & End of charging time for EV $k$, final charge energy for EV $k$ \\[2pt]
		$A\boldsymbol{s} + \boldsymbol{a}, F\boldsymbol{s} + \boldsymbol{f}, G\boldsymbol{s} + \boldsymbol{g}$ & linear models for mapping from node power injections to voltage magnitudes, transformer real power, and transformer reactive power \\[2pt]
		$\boldsymbol{\mu}, \lambda$ & cost of electricity, cost function component weight \\[2pt]
		\hline
	\end{tabular*}
\end{table*} 


\noindent{\bf Local control schemes}.  We consider two local control schemes. In both schemes, we assume that the controller at each consumer node is operating its DERs autonomously with the goal of minimizing the consumer's total electricity cost.

The first local controller, which we use throughout the study, is a simple heuristic that does not presume knowledge or forecasts of any variable values, except for the end of the current EV charging window and the percentage of thermal load that is flexible, which may be specified by the consumer.

In each time step, the controller checks if there is excess solar PV generation after meeting its node's uncontrollable demand. If there is excess PV generation it first uses it to charge an EV, assuming there is an EV to be charged, then uses it to power the flexible load, if any, up to the amount of flexible load demanded during the peak price period. Finally, any excess is used to charge the stationary storage unit, assuming there is such a system and the storage is not full. Any remaining excess PV generation beyond these three uses is fed back into the grid.

If there is no excess solar PV generation, the controller calculates the future time periods in which the EV (if exists) can charge at the lowest cost within the given charging window. The same procedure is applied to the stationary storage with the condition that it is fully charged just prior to the peak price period. If the current time step is within the lowest cost periods in the calculated schedules, the controller charges both the EV and storage. The controller reduces the power consumption of the flexible loads during the peak price period by as much as possible. After removing all flexible load from the peak price period, any remaining flexible load during the nearest partial peak load period is also removed. The percentage of load removed is upper bounded by the percentage of thermal load that is flexible, $\phi$. The same amount of energy removed is added back equally across 1 hour before and after the energy is reduced. Finally, the controller discharges the storage as much as needed or available to reduce demand during the peak price period.

While no local controller can directly reduce transformer overloading or voltage violations, the above local controller indirectly reduces them by trying to shift the load to non-peak time periods. One might then ask: how much can the performance of this controller improve both in terms of cost reduction and reliability improvement. To answer this question we consider the following perfect foresight local controller which assumes that each consumer node knows its load, including flexible load, solar PV generation, and EV charging windows over the entire 1-year simulation horizon in advance. The control actions for this scheme are then determined for each node $i\in [1:N_\mathrm{C}]$ by solving the optimization problem:
\vspace{-3pt}
\begin{subequations} \label{Localopt_op}
	\begin{alignat}{3}
 	&\underset{c_i, d_i, Q_i}{\text{minimize:}} \quad  L_{1,i}  + \lambda_2 L_{2,i}  \nonumber \\
    &	\text{subject to:} \nonumber \\  & \quad 0 \le \boldsymbol{c}_i(t) \le \boldsymbol{c}_i^{\mathrm{max}}, \label{cmax} \\
    &\quad 0 \le \boldsymbol{d}_i(t) \le \boldsymbol{d}_i^{\mathrm{max}}, \label{dmax} \\
    &\quad 0 \le \boldsymbol{u}_i(t) \le \boldsymbol{u}^{\mathrm{max}},  \label{umax} \\
    &\quad \boldsymbol{Q}_i(t) = \gamma_l \boldsymbol{Q}_i(t-1) + \gamma_c \boldsymbol{c}(t) - \gamma_d \boldsymbol{d}_i(t), \label{charge} \\
    &\quad \boldsymbol{Q}_i^{\mathrm{min}} \le \boldsymbol{Q}_i(t) \le \boldsymbol{Q}_i^{\mathrm{max}}, \label{qmax} \\
    &\quad Q_{ik}(t^{\mathrm{end}}_k) = Q_{ik}^{\mathrm{final}} \hspace{1em} \text{ for } k \in \mathcal{E}_i, \label{ev_charge} \\
    &\quad \sum_{t=1}^{T_d} u_i(t) = \sum_{t=1}^{T_d} u_i^{\mathrm{base}}(t), \label{thermal_energy} \\
    &\quad \frac{1}{2} \sum_{t=1}^{T_d} \lvert u_i(t) - u_i^{\mathrm{base}}(t) \rvert \le \phi_i \sum_{t=1}^{T_d} u_i^{\mathrm{base}}(t), \label{thermal_phi}
	\end{alignat}
\end{subequations} 
\vspace{-4pt}
where
\vspace{-3pt}
\begin{equation}
 L_{1i} =    \sum_{t=1}^{T} \mu_{i}(t) \Big[ p_i(t) + u_i(t) + \sum_{k=1}^{K_i}(c_{ik}(t) - d_{ik}(t)) \Big]_+  \label{energy_cost}
\end{equation}
is the total cost of electricity at the node, and
\begin{equation}
 L_{2i} =  \sum_{t=1}^{T} \sum_{k=1}^{K_i} (c_{ik}(t) + d_{ik}(t))^2 \label{battery_penalty}
\end{equation}
is a battery operating cost that aims to reduce battery aging. The constraints \eqref{cmax}, \eqref{dmax}, and \eqref{umax} limit the power each storage, EV charger, or flexible load can charge or discharge, \eqref{charge} models the storage charging dynamics, and \eqref{ev_charge} models the constraint that each EV must be fully charged within the given window. We are able to combine the EV charging and storage constraints by setting $d_{ik}^{\mathrm{max}} = 0$ to prevent vehicle to grid discharging and $c_{ik}^{\mathrm{max}} = 0$ when outside of the given charging window for each EV $k$. We assume EV charging has the same efficiency as stationary storage charging. The constraint \eqref{thermal_energy} ensures that the total daily energy before and after the optimization are the same and the constraint \eqref{thermal_phi} ensures that no more than $\phi$ of the thermal load is adjusted. The one half fraction prevents double counting both the curtailed and the compensated energy when balancing the energy consumption of the flexible thermal load. Note that the local controller used throughout the study effectively solves the optimization problem in~\eqref{Localopt_op} for a single time step $t$ without the $L_{2i}$ term in the objective.

In the Supplement we compare this perfect foresight local controller to the heuristic local controller we use in the study and discuss how costs are equivalent with the control schemes and reliability does not improve.
\smallskip

\noindent{\bf Centralized DER control scheme}. We consider a perfect foresight fully centralized DER control scheme with 2-day overlapping windows. The optimizer for each 2-day window has full knowledge of the load profile, solar generation, and EV charging events during these two days. The controller uses the optimization results for the first day and the process is repeated for the following 2-day window. This controller should provide a lower bound on the number of violations of any implementable scheme, since (i) an implementable scheme does not have any knowledge of the future, (ii) the operation of EVs and thermal loads are performed within a single day as the power cannot be deferred by more than one day, and (iii) the state of charge of the storage can reach all possible states within a few hours and has a cyclical nature due to the cyclical nature of solar PV power, loads, and the TOU tariffs. The main computational bottlenecks in simulating this scheme is the non-convexity of its optimization problems, the nonlinearity of the AC power flow equations for a 3-phase network, the large size of some of the networks, and the large number of DERs, especially flexible load and EV charging events. 


To address the non-convexity and non-linearity of the AC power flow equations, we use the following data-driven linear models of the AC power flow equations for each 3-phase network; see~\cite{PF_linear_survey, LinearPF_LSQ}. 
\begin{subequations} \label{eq:lin}
\begin{align}
	\boldsymbol{v}(t) &= A\boldsymbol{s}(t) + \boldsymbol{a}, \label{eq:lin_v}\\
	\boldsymbol{\tau}(t) &= (F\boldsymbol{s}(t) + \boldsymbol{f})^2 + (G \boldsymbol{s}(t) + \boldsymbol{g})^2 \label{eq:lin_t},
\end{align}
\end{subequations}
where for node $i \in [1:N]$,
 \[
 s_i(t) = \left[\begin{array}{c} p_i(t) + u_i(t) + \sum_{k=0}^{K_i}(c_{ik}(t) - d_{ik}(t) \\[3pt]
q_i(t)\end{array}\right] 
\] 
is the power at time $t\in[1:T]$.
This approximation makes the optimization problem convex and computationally tractable and have been shown to be quite accurate. 

The linear model coefficients are trained via least squares regression. For each centralized controller scenario, the training data consists of the nodal voltage / transformer apparent power and power injection data from the peak month of the corresponding local controller simulation.

In addition to the costs of electricity~\eqref{energy_cost} and storage~\eqref{battery_penalty}, the cost function for the centralized scheme include a component for voltage violations 
\begin{equation*}
L_3 =    \sum_{t=1}^{T} \sum_{i=1}^{N} \left( \lbrack v_i(t) - v_i^{\mathrm{max}} \rbrack_+ + \lbrack v_i^{\mathrm{min}} - v_i(t) \rbrack_+ \right)^2, \label{v_penalty}
\end{equation*}
and a component for transformer overloading
\begin{equation}
L_4 =  \sum_{t=1}^{T} \sum_{(i,j) \in \mathcal{T}} \big( \lbrack \tau_{ij}(t) - {\tau_{ij}^{\mathrm{max}}} \rbrack_+ \big)^2.
    \label{t_penalty}
\end{equation}
The limits in these two cost functions are tightened by 5\% compared to the metric to account for linear model inaccuracies and to discourage even the smallest of violations.

The actions of the centralized DER controller are obtained by solving the following optimization problem for overlapping 2-day windows over the 1-year simulation horizon.  
\begin{subequations} \label{eq:central_opt_grid}
	\begin{alignat}{2}
		&\underset{\boldsymbol{c}, \boldsymbol{d}, \boldsymbol{Q}, \boldsymbol{v}, \boldsymbol{\tau}} {\text{minimize:}} \quad \sum_{i=1}^{N_\mathrm{C}} \big( L_{1i} + \lambda_2 L_{2i} \big) + \lambda_3 L_3 + \lambda_4 L_4 \nonumber \\
		&\text{subject to:} \quad \eqref{eq:lin_v}, \eqref{eq:lin_t}, \nonumber \\
		&\qquad \eqref{cmax},\eqref{dmax},\eqref{umax}, \nonumber \\
        &\qquad  \eqref{charge}, \eqref{qmax}, \eqref{ev_charge}, \eqref{thermal_energy},\eqref{thermal_phi} \hspace{1em} \text{ for } i \in [1:N_\mathrm{C}].  \nonumber
	\end{alignat} 
\end{subequations} 
Note that the cost function for this optimization problem is a weighted sum of the electricity, storage operation, voltage violations and transformer overloading costs. The weights were selected to emphasize the priority of each cost component. The unweighted cost (energy cost) is determined by the cost of electricity which has a maximum value of 0.125. The grid reliability objectives have maximum priority so we gave it a weight of 125. The battery operating cost has the lowest priority so has a weight of 0.001. These weights were determined by performing several experiments to show that further emphasis on reliability does not yield different results.
We use the linearized power flow approximations to make the problem computationally feasible.
The constraints are for the storage operation, EV charging, and flexible load.
\smallskip

\noindent{\bf Power flow simulation} \noindent{(Figure~\ref{fig:block_diagram} 6)}. To obtain the voltage magnitudes and transformer apparent powers from the net power injections at each node in the network scenario, we use the OpenDSS quasi-static power flow simulator~\cite{montenegro2012real}. 
Following~\cite{hosting_capacity}, the metric we use for steady-state voltage declares a violation at a node if its voltage magnitude exceeds the specifications in the ANSI C84.1 standard, which represent a deviation of $\pm 5\%$ from the nominal voltage magnitude in one or more 15-minute time steps. The metric for transformer thermal overloading declares a violation at a transformer if its average apparent power is greater than 120\% of its rated capacity over one or more 2-hour windows.

\subsection*{Statistical methods}

\noindent{(Figure~\ref{fig:block_diagram} 8)}. In order to calculate the averages and standard deviations reported in all figures with average results across all networks, we first calculate the sample mean and standard deviation across all 16 scenarios for each of the 11 networks, leading to 11 means and 11 standard deviations. We then compute the average and standard deviation, as the square root of the average sum of the variances, across all 11 networks.

\subsection*{Software implementation} 

To obtain the results reported in this paper, we ran 27 simulations for each of the 11 networks. Each simulation involved the generation of 16, 1-year scenarios, each at a 15 minute resolution. The main computational bottlenecks in our simulations are performing very large numbers of local control operations (around 14 million) to determine DER power injections and solving the optimization problems for the centralized controllers. To address the first challenge, since the tasks are \emph{embarrassingly parallel}, we ran our simulations on a large number of processor cores. We addressed the second challenge by making the approximations discussed in the control schemes section.
The largest simulation in our study is the Oakland network in the year 2050, which has 10073 nodes, 1789 EVs, 1494 storage units, resulting in the optimization problem having 8 million variables for each 2-day window. The optimization is solved 365 times sequentially, updating the inputs for the next day. Even after addressing the aforementioned computational bottlenecks, the computation time for solving this optimization problem takes over 64 minutes on an Intel Xeon E5-2640 V4 2.6GHz 8-Core processor.
\backmatter


\section*{Funding}
The work presented in this paper was supported under ARPA-E award DE-AR0000697 and US DOE award DE-OE0000919.

\section*{Authors' contributions}
T.N., A.E.G., and R.R. conceptualized the study. T.N. performed the data curation, formal analysis, collection of resources, investigation, creation of methodology and software, validation, and visualization. T.N. and A.E.G. wrote the original draft. A.E.G. provided supervision and project administration. R.R. and A.E.G. provided review, and funding acquisition.

\section*{Declaration of interests}
The authors declare no competing interests

\section*{Inclusion and diversity}
We support inclusive, diverse, and equitable conduct of research.



\bibliography{sn-bibliography}

\begin{thebibliography}{10}
\expandafter\ifx\csname url\endcsname\relax
  \def\url#1{\burl{#1}}\fi
\expandafter\ifx\csname urlprefix\endcsname\relax\def\urlprefix{URL }\fi
\providecommand{\bibinfo}[2]{#2}
\providecommand{\eprint}[2][]{\url{#2}}
\providecommand{\doi}[1]{\url{https://doi.org/#1}}
\bibcommenthead

\bibitem{mitnick2015changing}
\bibinfo{author}{Mitnick, S.} \& \bibinfo{author}{America, B.~E.}
\newblock \bibinfo{title}{Changing uses of the electric grid: reliability
  challenges and concerns}.
\newblock \emph{\bibinfo{journal}{Electric Markets Research Foundation}}
  (\bibinfo{year}{2015}) .

\bibitem{reliability_survey}
\bibinfo{author}{Sultan, V.} \& \bibinfo{author}{Hilton, B.}
\newblock \bibinfo{title}{Electric grid reliability research}.
\newblock \emph{\bibinfo{journal}{Energy Informatics}}
  \textbf{\bibinfo{volume}{2}}~(1), \bibinfo{pages}{3} (\bibinfo{year}{2019}).
\newblock \doi{10.1186/s42162-019-0069-z} .

\bibitem{PV_regulators_journal}
\bibinfo{author}{ElNozahy, M.~S.} \& \bibinfo{author}{Salama, M. M.~A.}
\newblock \bibinfo{title}{Technical impacts of grid-connected photovoltaic
  systems on electrical networks - a review}.
\newblock \emph{\bibinfo{journal}{Journal of Renewable and Sustainable Energy}}
  \textbf{\bibinfo{volume}{5}} (\bibinfo{year}{2013}) .

\bibitem{EV_impacts_nature}
\bibinfo{author}{Muratori, M.}
\newblock \bibinfo{title}{Impact of uncoordinated plug-in electric vehicle
  charging on residential power demand}.
\newblock \emph{\bibinfo{journal}{Nature Energy}}
  \textbf{\bibinfo{volume}{3}}~(3), \bibinfo{pages}{193--201}
  (\bibinfo{year}{2018}).
\newblock \doi{10.1038/s41560-017-0074-z} .

\bibitem{cost_upgrades_NREL}
\bibinfo{author}{Horowitz, K. A.~W.}, \bibinfo{author}{Ding, F.},
  \bibinfo{author}{Mather, B.} \& \bibinfo{author}{Palmintier, B.}
\newblock \bibinfo{title}{The cost of distribution system upgrades to
  accommodate increasing penetrations of distributed photovoltaic systems on
  real feeders in the united states}.
\newblock \bibinfo{type}{Tech. Rep.} \bibinfo{number}{NREL/TP-6A20-70710},
  \bibinfo{institution}{National Renewable Energy Laboratory},
  \bibinfo{address}{Golden, CO} (\bibinfo{year}{2018}).
\newblock \urlprefix\url{https://www.nrel.gov/docs/fy18osti/70710.pdf}.

\bibitem{upgrade_cost_PV}
\bibinfo{author}{Horowitz, K.~A.}, \bibinfo{author}{Palmintier, B.},
  \bibinfo{author}{Mather, B.} \& \bibinfo{author}{Denholm, P.}
\newblock \bibinfo{title}{Distribution system costs associated with the
  deployment of photovoltaic systems}.
\newblock \emph{\bibinfo{journal}{Renewable and Sustainable Energy Reviews}}
  \textbf{\bibinfo{volume}{90}}, \bibinfo{pages}{420--433}
  (\bibinfo{year}{2018}).
\newblock
  \urlprefix\url{https://www.sciencedirect.com/science/article/pii/S1364032118301734}.
\newblock \doi{https://doi.org/10.1016/j.rser.2018.03.080} .

\bibitem{national2017enhancing}
\bibinfo{author}{National Academies~of Sciences, .~M., Engineering}.
\newblock \emph{\bibinfo{title}{Enhancing the Resilience of the Nation's
  Electricity System}}  (\bibinfo{publisher}{The National Academies Press},
  \bibinfo{address}{Washington, DC}, \bibinfo{year}{2017}).
\newblock
  \urlprefix\url{https://nap.nationalacademies.org/catalog/24836/enhancing-the-resilience-of-the-nations-/electricity-system}.

\bibitem{price_transactive_nature}
\bibinfo{author}{Wang, N.}
\newblock \bibinfo{title}{Transactive control for connected homes and
  neighbourhoods}.
\newblock \emph{\bibinfo{journal}{Nature Energy}}
  \textbf{\bibinfo{volume}{3}}~(11), \bibinfo{pages}{907--909}
  (\bibinfo{year}{2018}).
\newblock \doi{10.1038/s41560-018-0257-2} .

\bibitem{price_dynamic_residential}
\bibinfo{author}{Borenstein, S.}
\newblock \bibinfo{title}{Effective and equitable adoption of opt-in
  residential dynamic electricity pricing}.
\newblock \emph{\bibinfo{journal}{Review of Industrial Organization}}
  \textbf{\bibinfo{volume}{42}}~(2), \bibinfo{pages}{127--160}
  (\bibinfo{year}{2013}).
\newblock \doi{10.1007/s11151-012-9367-3} .

\bibitem{price_DLMP}
\bibinfo{author}{Andrianesis, P.} \& \bibinfo{author}{Caramanis, M.}
\newblock \bibinfo{title}{Distribution network marginal costs: Enhanced ac opf
  including transformer degradation}.
\newblock \emph{\bibinfo{journal}{IEEE Transactions on Smart Grid}}
  \textbf{\bibinfo{volume}{11}}~(5), \bibinfo{pages}{3910--3920}
  (\bibinfo{year}{2020}).
\newblock \doi{10.1109/TSG.2020.2980538} .

\bibitem{grid_models_NREL}
\bibinfo{author}{Palmintier, B.} \emph{et~al.}
\newblock \bibinfo{title}{Experiences developing large-scale synthetic
  u.s.-style distribution test systems}.
\newblock \emph{\bibinfo{journal}{Electric Power Systems Research}}
  \textbf{\bibinfo{volume}{190}}, \bibinfo{pages}{106665}
  (\bibinfo{year}{2021}).
\newblock
  \urlprefix\url{https://www.sciencedirect.com/science/article/pii/S0378779620304685}.
\newblock \doi{https://doi.org/10.1016/j.epsr.2020.106665} .

\bibitem{iowa}
\bibinfo{author}{Bu, F.}, \bibinfo{author}{Yuan, Y.}, \bibinfo{author}{Wang,
  Z.}, \bibinfo{author}{Dehghanpour, K.} \& \bibinfo{author}{Kimber, A.}
\newblock \bibinfo{title}{A time-series distribution test system based on real
  utility data}.
\newblock \emph{\bibinfo{journal}{2019 North American Power Symposium (NAPS)}}
  \bibinfo{pages}{1--6} (\bibinfo{year}{2019}) .

\bibitem{test_feeders_IEEE}
\bibinfo{author}{Schneider, K.~P.} \emph{et~al.}
\newblock \bibinfo{title}{Analytic considerations and design basis for the ieee
  distribution test feeders}.
\newblock \emph{\bibinfo{journal}{IEEE Transactions on Power Systems}}
  \textbf{\bibinfo{volume}{PP}}~(99), \bibinfo{pages}{1--1}
  (\bibinfo{year}{2017}) .

\bibitem{EPRI_J1}
\bibinfo{title}{Electric power research institute (epri) feeder j1}
  (\bibinfo{year}{2022}).
\newblock \urlprefix\url{https://sourceforge.net/p/electricdss/
  code/HEAD/tree/trunk/Distrib/ EPRITestCircuits/epri\_dpv/J1/}.

\bibitem{restock_nrel}
\bibinfo{author}{Wilson, E.}, \bibinfo{author}{Christensen, C.},
  \bibinfo{author}{Horowitz, S.}, \bibinfo{author}{Robertson, J.} \&
  \bibinfo{author}{Maguire, J.}
\newblock \bibinfo{title}{Energy efficiency potential in the u.s. single-family
  housing stock}.
\newblock \bibinfo{type}{Tech. Rep.} \bibinfo{number}{NREL/TP-5500-68670},
  \bibinfo{institution}{National Renewable Energy Laboratory},
  \bibinfo{address}{Golden, CO} (\bibinfo{year}{December 2017}).
\newblock \urlprefix\url{https://www.nrel.gov/docs/fy18osti/68670.pdf}.

\bibitem{NREL_stock_data}
\bibinfo{author}{Wilson, E. J.~H.} \emph{et~al.}
\newblock \bibinfo{title}{End-use load profiles for the u.s. building stock:
  Methodology and results of model calibration, validation, and uncertainty
  quantification}  (\bibinfo{year}{2022}).
\newblock \urlprefix\url{https://www.osti.gov/biblio/1854582}.
\newblock \doi{10.2172/1854582} .

\bibitem{NREL_EV_adoption}
\bibinfo{author}{Trieu, M.} \emph{et~al.}
\newblock \bibinfo{title}{Electrification futures study: Scenarios of electric
  technology adoption and power consumption for the united states}.
\newblock \bibinfo{type}{Tech. Rep.} \bibinfo{number}{NREL/TP-6A20-71500},
  \bibinfo{institution}{NREL}, \bibinfo{address}{Golden, CO, USA}
  (\bibinfo{year}{2018}).
\newblock \urlprefix\url{https://www.nrel.gov/docs/fy18osti/71500.pdf}.

\bibitem{NREL_storage}
\bibinfo{author}{Prasanna, A.}, \bibinfo{author}{McCabe, K.},
  \bibinfo{author}{Sigrin, B.} \& \bibinfo{author}{Blair, N.}
\newblock \bibinfo{title}{Storage futures study: Distributed solar and storage
  outlook: Methodology and scenarios.}
\newblock \bibinfo{type}{Tech. Rep.} \bibinfo{number}{NREL/TP-7A40-79790},
  \bibinfo{institution}{NREL}, \bibinfo{address}{Golden, CO, USA}
  (\bibinfo{year}{2021}).
\newblock \urlprefix\url{https://www.nrel.gov/docs/fy21osti/79790.pdf}.

\bibitem{flex_load_NREL}
\bibinfo{author}{Jadun, P.} \emph{et~al.}
\newblock \bibinfo{title}{Electrification futures study flexible load
  profiles}.
\newblock \bibinfo{type}{Tech. Rep.}, \bibinfo{institution}{National Renewable
  Energy Laboratory}, \bibinfo{address}{Golden, CO} (\bibinfo{year}{2020}).
\newblock \urlprefix\url{https://data.nrel.gov/submissions/127}.

\bibitem{pge_tariffs}
\bibinfo{title}{Pacific gas and electric: Tariffs} (\bibinfo{year}{2022}).
\newblock \bibinfo{note}{\url{https://www.pge.com/tariffs/index.page}}.

\bibitem{hosting_capacity}
\bibinfo{author}{Jain, A.~K.} \emph{et~al.}
\newblock \bibinfo{title}{Quasi-static time-series pv hosting capacity
  methodology and metrics}.
\newblock \emph{\bibinfo{journal}{2019 IEEE Power \& Energy Society Innovative
  Smart Grid Technologies Conference (ISGT)}} \bibinfo{pages}{1--5}
  (\bibinfo{year}{2019}).
\newblock \doi{10.1109/ISGT.2019.8791569} .

\bibitem{capacity_analysis_survey}
\bibinfo{author}{Ismael, S.~M.}, \bibinfo{author}{Aleem, A.},
  \bibinfo{author}{Abdelaziz, A.~Y.} \& \bibinfo{author}{Zobaa, A.~F.}
\newblock \bibinfo{title}{State-of-the-art of hosting capacity in modern power
  systems with distributed generation}.
\newblock \emph{\bibinfo{journal}{Renewable Energy}}
  \textbf{\bibinfo{volume}{130}}, \bibinfo{pages}{1002--1020}
  (\bibinfo{year}{2019}).
\newblock
  \urlprefix\url{https://www.sciencedirect.com/science/article/pii/S0960148118307936}.
\newblock \doi{https://doi.org/10.1016/j.renene.2018.07.008} .

\bibitem{capacity_analysis_dynamic_NREL}
\bibinfo{author}{Jain, A.~K.} \emph{et~al.}
\newblock \bibinfo{title}{Dynamic hosting capacity analysis for distributed
  photovoltaic resources—framework and case study}.
\newblock \emph{\bibinfo{journal}{Applied Energy}}
  \textbf{\bibinfo{volume}{280}}, \bibinfo{pages}{115633}
  (\bibinfo{year}{2020}).
\newblock
  \urlprefix\url{https://www.sciencedirect.com/science/article/pii/S0306261920311351}.
\newblock \doi{https://doi.org/10.1016/j.apenergy.2020.115633} .

\bibitem{capacity_analysis_EPRI}
\bibinfo{author}{Rylander, M.}, \bibinfo{author}{Smith, J.} \&
  \bibinfo{author}{Sunderman, W.}
\newblock \bibinfo{title}{Streamlined method for determining distribution
  system hosting capacity}.
\newblock \emph{\bibinfo{journal}{IEEE Transactions on Industry Applications}}
  \textbf{\bibinfo{volume}{52}}~(1), \bibinfo{pages}{105--111}
  (\bibinfo{year}{2016}).
\newblock \doi{10.1109/TIA.2015.2472357} .

\bibitem{capacity_analysis_storage_alg}
\bibinfo{author}{Hashemi, S.} \& \bibinfo{author}{Østergaard, J.}
\newblock \bibinfo{title}{Efficient control of energy storage for increasing
  the pv hosting capacity of lv grids}.
\newblock \emph{\bibinfo{journal}{IEEE Transactions on Smart Grid}}
  \textbf{\bibinfo{volume}{9}}~(3), \bibinfo{pages}{2295--2303}
  (\bibinfo{year}{2018}).
\newblock \doi{10.1109/TSG.2016.2609892} .

\bibitem{capacity_analysis_storage_swiss}
\bibinfo{author}{Gupta, R.}, \bibinfo{author}{Sossan, F.} \&
  \bibinfo{author}{Paolone, M.}
\newblock \bibinfo{title}{Countrywide pv hosting capacity and energy storage
  requirements for distribution networks: The case of switzerland}.
\newblock \emph{\bibinfo{journal}{Applied Energy}}
  \textbf{\bibinfo{volume}{281}}, \bibinfo{pages}{116010}
  (\bibinfo{year}{2021}).
\newblock
  \urlprefix\url{https://www.sciencedirect.com/science/article/pii/S0306261920314537}.
\newblock \doi{https://doi.org/10.1016/j.apenergy.2020.116010} .

\bibitem{capacity_analysis_EVs_NREL}
\bibinfo{author}{Paudyal, P.}, \bibinfo{author}{Ghosh, S.},
  \bibinfo{author}{Veda, S.}, \bibinfo{author}{Tiwari, D.} \&
  \bibinfo{author}{Desai, J.}
\newblock \bibinfo{title}{Ev hosting capacity analysis on distribution grids:
  Preprint}  (\bibinfo{year}{2021}).
\newblock \urlprefix\url{https://www.osti.gov/biblio/1818876} .

\bibitem{tariff_house_nature}
\bibinfo{author}{Azarova, V.}, \bibinfo{author}{Engel, D.},
  \bibinfo{author}{Ferner, C.}, \bibinfo{author}{Kollmann, A.} \&
  \bibinfo{author}{Reichl, J.}
\newblock \bibinfo{title}{Exploring the impact of network tariffs on household
  electricity expenditures using load profiles and socio-economic
  characteristics}.
\newblock \emph{\bibinfo{journal}{Nature Energy}}
  \textbf{\bibinfo{volume}{3}}~(4), \bibinfo{pages}{317--325}
  (\bibinfo{year}{2018}).
\newblock \doi{10.1038/s41560-018-0105-4} .

\bibitem{VGI_nature}
\bibinfo{author}{Wolinetz, M.}, \bibinfo{author}{Axsen, J.},
  \bibinfo{author}{Peters, J.} \& \bibinfo{author}{Crawford, C.}
\newblock \bibinfo{title}{Simulating the value of electric-vehicle--grid
  integration using a behaviourally realistic model}.
\newblock \emph{\bibinfo{journal}{Nature Energy}}
  \textbf{\bibinfo{volume}{3}}~(2), \bibinfo{pages}{132--139}
  (\bibinfo{year}{2018}).
\newblock \doi{10.1038/s41560-017-0077-9} .

\bibitem{DER_LCOE_ram_nature}
\bibinfo{author}{Jain, R.~K.}, \bibinfo{author}{Qin, J.} \&
  \bibinfo{author}{Rajagopal, R.}
\newblock \bibinfo{title}{Data-driven planning of distributed energy resources
  amidst socio-technical complexities}.
\newblock \emph{\bibinfo{journal}{Nature Energy}}
  \textbf{\bibinfo{volume}{2}}~(8), \bibinfo{pages}{17112}
  (\bibinfo{year}{2017}).
\newblock \doi{10.1038/nenergy.2017.112} .

\bibitem{grid_infrastructure_inequity_nature}
\bibinfo{author}{Brockway, A.~M.}, \bibinfo{author}{Conde, J.} \&
  \bibinfo{author}{Callaway, D.}
\newblock \bibinfo{title}{Inequitable access to distributed energy resources
  due to grid infrastructure limits in california}.
\newblock \emph{\bibinfo{journal}{Nature Energy}}
  \textbf{\bibinfo{volume}{6}}~(9), \bibinfo{pages}{892--903}
  (\bibinfo{year}{2021}).
\newblock \doi{10.1038/s41560-021-00887-6} .

\bibitem{solar_compete_nature}
\bibinfo{author}{Borenstein, S.}
\newblock \bibinfo{title}{It's time for rooftop solar to compete with other
  renewables}.
\newblock \emph{\bibinfo{journal}{Nature Energy}}
  \textbf{\bibinfo{volume}{7}}~(4), \bibinfo{pages}{298--298}
  (\bibinfo{year}{2022}).
\newblock \doi{10.1038/s41560-022-01015-8} .

\bibitem{data_opf_guo}
\bibinfo{author}{Guo, Y.}, \bibinfo{author}{Baker, K.},
  \bibinfo{author}{Dall'Anese, E.}, \bibinfo{author}{Hu, Z.} \&
  \bibinfo{author}{Summers, T.}
\newblock \bibinfo{title}{Data-based distributionally robust stochastic optimal
  power flow—part i: Methodologies}.
\newblock \emph{\bibinfo{journal}{IEEE Transactions on Power Systems}}
  \textbf{\bibinfo{volume}{34}}, \bibinfo{pages}{1483--1492}
  (\bibinfo{year}{2018}) .

\bibitem{DMPC}
\bibinfo{author}{Braun, P.}, \bibinfo{author}{Grune, L.},
  \bibinfo{author}{Kellett, C.~M.}, \bibinfo{author}{Weller, S.~R.} \&
  \bibinfo{author}{Worthmann, K.}
\newblock \bibinfo{title}{Predictive control of a smart grid: A distributed
  optimization algorithm with centralized performance properties}.
\newblock \emph{\bibinfo{journal}{IEEE 54th Annual Conference on Decision and
  Control}}  (\bibinfo{year}{2015}) .

\bibitem{boyd_messaging}
\bibinfo{author}{Kraning, M.}, \bibinfo{author}{Chu, E.},
  \bibinfo{author}{Lavaei, J.} \& \bibinfo{author}{Boyd, S.~P.}
\newblock \bibinfo{title}{Dynamic network energy management via proximal
  message passing.}
\newblock \emph{\bibinfo{journal}{Foundations and Trends in Optimization}}
  \textbf{\bibinfo{volume}{1}}~(2), \bibinfo{pages}{73--126}
  (\bibinfo{year}{2014}) .

\bibitem{junjie}
\bibinfo{author}{Qin, J.}, \bibinfo{author}{Chow, Y.}, \bibinfo{author}{Yang,
  J.} \& \bibinfo{author}{Rajagopal, R.}
\newblock \bibinfo{title}{Distributed online modified greedy algorithm for
  networked storage operation under uncertainty}.
\newblock \emph{\bibinfo{journal}{Smart Grid, IEEE Transactions on}}
  \textbf{\bibinfo{volume}{7}}~(2), \bibinfo{pages}{1106--1118}
  (\bibinfo{year}{2015}) .

\bibitem{consensus}
\bibinfo{author}{Wang, Y.}, \bibinfo{author}{Tan, K.~T.},
  \bibinfo{author}{Peng, X.~Y.} \& \bibinfo{author}{So, P.~L.}
\newblock \bibinfo{title}{Coordinated control of distributed energy storage
  systems for voltage regulation in distribution networks}.
\newblock \emph{\bibinfo{journal}{IEEE Trans on Power Delivery}}
  \textbf{\bibinfo{volume}{31}}~(3) (\bibinfo{year}{2016}) .

\bibitem{P2P_VPP_nature}
\bibinfo{author}{Morstyn, T.}, \bibinfo{author}{Farrell, N.},
  \bibinfo{author}{Darby, S.~J.} \& \bibinfo{author}{McCulloch, M.~D.}
\newblock \bibinfo{title}{Using peer-to-peer energy-trading platforms to
  incentivize prosumers to form federated power plants}.
\newblock \emph{\bibinfo{journal}{Nature Energy}}
  \textbf{\bibinfo{volume}{3}}~(2), \bibinfo{pages}{94--101}
  (\bibinfo{year}{2018}).
\newblock \doi{10.1038/s41560-017-0075-y} .

\bibitem{kyle_anderson:2017}
\bibinfo{author}{Anderson, K.}, \bibinfo{author}{Rajagopal, R.} \&
  \bibinfo{author}{El~Gamal, A.}
\newblock \bibinfo{title}{Coordination of distributed storage under temporal
  and spatial data asymmetry}.
\newblock \emph{\bibinfo{journal}{IEEE Trans. on Smart Grid}}
  \textbf{\bibinfo{volume}{PP}}~(99) (\bibinfo{year}{2017}) .

\bibitem{Navidi_smartgridcomm}
\bibinfo{author}{Navidi, T.}, \bibinfo{author}{Leblanc, C.},
  \bibinfo{author}{El~Gamal, A.} \& \bibinfo{author}{Rajagopal, R.}
\newblock \bibinfo{title}{Der information unaware coordination via day-ahead
  dynamic power bounds}.
\newblock \emph{\bibinfo{journal}{IEEE International Conference on
  Communications, Control, and Computing Technologies for Smart Grids
  (SmartGridComm)}} \bibinfo{pages}{1--6} (\bibinfo{year}{2020}).
\newblock \doi{10.1109/SmartGridComm47815.2020.9302955} .

\bibitem{setpoint_track_Bernstein}
\bibinfo{author}{Bernstein, A.}, \bibinfo{author}{Reyes, L.},
  \bibinfo{author}{Le~Boudec, J.-Y.} \& \bibinfo{author}{Paolone, M.}
\newblock \bibinfo{title}{A composable method for real-time control of active
  distribution networks with explicit power setpoints}.
\newblock \emph{\bibinfo{journal}{Electric Power Systems Research}}
  \textbf{\bibinfo{volume}{125}} (\bibinfo{year}{2014}).
\newblock \doi{10.1016/j.epsr.2015.03.023} .

\bibitem{PacketizedEnergy}
\bibinfo{author}{Espinosa, L. A.~D.}, \bibinfo{author}{Almassalkhi, M.},
  \bibinfo{author}{Hines, P.} \& \bibinfo{author}{Frolik, J.}
\newblock \bibinfo{title}{System properties of packetized energy management for
  aggregated diverse resources}.
\newblock \emph{\bibinfo{journal}{Power Systems Computation Conference (PSCC)}}
   (\bibinfo{year}{2018}) .

\bibitem{data_ML_local}
\bibinfo{author}{Karagiannopoulos, S.}, \bibinfo{author}{Aristidou, P.} \&
  \bibinfo{author}{Hug, G.}
\newblock \bibinfo{title}{Data-driven local control design for active
  distribution grids using off-line optimal power flow and machine learning
  techniques}.
\newblock \emph{\bibinfo{journal}{IEEE Transactions on Smart Grid}}
  \textbf{\bibinfo{volume}{10}}, \bibinfo{pages}{6461--6471}
  (\bibinfo{year}{2019}) .

\bibitem{horowitz_derms}
\bibinfo{author}{Yao, Y.}, \bibinfo{author}{Ding, F.},
  \bibinfo{author}{Horowitz, K.} \& \bibinfo{author}{Jain, A.}
\newblock \bibinfo{title}{Coordinated inverter control to increase dynamic pv
  hosting capacity: A real-time optimal power flow approach}.
\newblock \emph{\bibinfo{journal}{IEEE Systems Journal}}
  \textbf{\bibinfo{volume}{16}}~(2), \bibinfo{pages}{1933--1944}
  (\bibinfo{year}{2022}).
\newblock \doi{10.1109/JSYST.2021.3071998} .

\bibitem{Xi_coop}
\bibinfo{author}{Xi, X.}, \bibinfo{author}{Sioshansi, R.} \&
  \bibinfo{author}{Marano, V.}
\newblock \bibinfo{title}{A stochastic dynamic programming model for
  co-optimization of distributed energy storage}.
\newblock \emph{\bibinfo{journal}{Energy Systems}}
  \textbf{\bibinfo{volume}{5}}~(3), \bibinfo{pages}{475--505}
  (\bibinfo{year}{2014}) .

\bibitem{kyle_coop}
\bibinfo{author}{Anderson, K.} \& \bibinfo{author}{El~Gamal, A.}
\newblock \bibinfo{title}{Co-optimizing the value of storage in energy and
  regulation service markets}.
\newblock \emph{\bibinfo{journal}{Energy Systems}} \textbf{\bibinfo{volume}{2}}
  (\bibinfo{year}{2017}) .

\bibitem{Zhang_coop}
\bibinfo{author}{Shi, Y.}, \bibinfo{author}{Xu, B.}, \bibinfo{author}{Wang, D.}
  \& \bibinfo{author}{Zhang, B.}
\newblock \bibinfo{title}{Using battery storage for peak shaving and frequency
  regulation: Joint optimization for superlinear gains}.
\newblock \emph{\bibinfo{journal}{IEEE Transactions on Power Systems}}
  \textbf{\bibinfo{volume}{PP}} (\bibinfo{year}{2017}).
\newblock \doi{10.1109/TPWRS.2017.2749512} .

\bibitem{NEC_coop}
\bibinfo{author}{Arabali, A.}, \bibinfo{author}{Asghari, B.} \&
  \bibinfo{author}{Sharma, R.}
\newblock \bibinfo{title}{A new co-optimization model for grid scale storage
  units in energy and frequency regulation markets}.
\newblock \emph{\bibinfo{journal}{IEEE Transmission and Distribution Conference
  and Exposition}}  (\bibinfo{year}{2016}) .

\bibitem{building_carbon}
\bibinfo{author}{Langevin, J.}, \bibinfo{author}{Harris, C.~B.} \&
  \bibinfo{author}{Reyna, J.~L.}
\newblock \bibinfo{title}{Assessing the potential to reduce u.s. building co2
  emissions 80\% by 2050}.
\newblock \emph{\bibinfo{journal}{Joule}} \textbf{\bibinfo{volume}{3}}~(10),
  \bibinfo{pages}{2403--2424} (\bibinfo{year}{2019}).
\newblock
  \urlprefix\url{https://www.sciencedirect.com/science/article/pii/S2542435119303575}.
\newblock \doi{https://doi.org/10.1016/j.joule.2019.07.013} .

\bibitem{building_resource}
\bibinfo{author}{Langevin, J.} \emph{et~al.}
\newblock \bibinfo{title}{Us building energy efficiency and flexibility as an
  electric grid resource}.
\newblock \emph{\bibinfo{journal}{Joule}} \textbf{\bibinfo{volume}{5}}~(8),
  \bibinfo{pages}{2102--2128} (\bibinfo{year}{2021}).
\newblock
  \urlprefix\url{https://www.sciencedirect.com/science/article/pii/S2542435121002907}.
\newblock \doi{https://doi.org/10.1016/j.joule.2021.06.002} .

\bibitem{building_resource_prev}
\bibinfo{author}{Jackson, R.}, \bibinfo{author}{Zhou, E.} \&
  \bibinfo{author}{Reyna, J.}
\newblock \bibinfo{title}{Building and grid system benefits of demand
  flexibility and energy efficiency}.
\newblock \emph{\bibinfo{journal}{Joule}} \textbf{\bibinfo{volume}{5}}~(8),
  \bibinfo{pages}{1927--1930} (\bibinfo{year}{2021}).
\newblock
  \urlprefix\url{https://www.sciencedirect.com/science/article/pii/S2542435121003573}.
\newblock \doi{https://doi.org/10.1016/j.joule.2021.08.001} .

\bibitem{space_heating}
\bibinfo{author}{Waite, M.} \& \bibinfo{author}{Modi, V.}
\newblock \bibinfo{title}{Electricity load implications of space heating
  decarbonization pathways}.
\newblock \emph{\bibinfo{journal}{Joule}} \textbf{\bibinfo{volume}{4}}~(2),
  \bibinfo{pages}{376--394} (\bibinfo{year}{2020}).
\newblock
  \urlprefix\url{https://www.sciencedirect.com/science/article/pii/S2542435119305781}.
\newblock \doi{https://doi.org/10.1016/j.joule.2019.11.011} .

\bibitem{tracking_thermal}
\bibinfo{author}{Pang, S.} \emph{et~al.}
\newblock \bibinfo{title}{Collaborative power tracking method of diversified
  thermal loads for optimal demand response: A milp-based decomposition
  algorithm}.
\newblock \emph{\bibinfo{journal}{Applied Energy}}
  \textbf{\bibinfo{volume}{327}}, \bibinfo{pages}{120006}
  (\bibinfo{year}{2022}).
\newblock
  \urlprefix\url{https://www.sciencedirect.com/science/article/pii/S0306261922012636}.
\newblock \doi{https://doi.org/10.1016/j.apenergy.2022.120006} .

\bibitem{exergy_replacement}
\bibinfo{author}{Ding, J.}, \bibinfo{author}{Gao, C.}, \bibinfo{author}{Song,
  M.}, \bibinfo{author}{Yan, X.} \& \bibinfo{author}{Chen, T.}
\newblock \bibinfo{title}{Bi-level optimal scheduling of virtual energy station
  based on equal exergy replacement mechanism}.
\newblock \emph{\bibinfo{journal}{Applied Energy}}
  \textbf{\bibinfo{volume}{327}}, \bibinfo{pages}{120055}
  (\bibinfo{year}{2022}).
\newblock
  \urlprefix\url{https://www.sciencedirect.com/science/article/pii/S0306261922013125}.
\newblock \doi{https://doi.org/10.1016/j.apenergy.2022.120055} .

\bibitem{BCG}
\bibinfo{author}{Sahoo, A.}, \bibinfo{author}{Mistry, K.} \&
  \bibinfo{author}{Baker, T.}
\newblock \bibinfo{title}{The costs of revving up the grid for electric
  vehicles}.
\newblock
  \bibinfo{howpublished}{\url{https://www.bcg.com/publications/2019/costs-revving-up-the-grid-for-electric-vehicles}}
  (\bibinfo{year}{2019}).
\newblock \bibinfo{note}{Accessed: 2023}.

\bibitem{EIA_grid_cost}
\bibinfo{author}{Aniti, L.} \& \bibinfo{author}{EIA}.
\newblock \bibinfo{title}{Major utilities’ spending on the electric
  distribution system continues to increase}.
\newblock
  \bibinfo{howpublished}{\url{https://www.eia.gov/todayinenergy/detail.php?id=48136}}
  (\bibinfo{year}{2021}).
\newblock \bibinfo{note}{Accessed: 2023}.

\bibitem{EV_charging_CEC}
\bibinfo{author}{Alexander, M.}, \bibinfo{author}{Crisostomo, N.},
  \bibinfo{author}{Krell, W.}, \bibinfo{author}{Lu, J.} \&
  \bibinfo{author}{Ramesh, R.}
\newblock \bibinfo{title}{Assembly bill 2127 electric vehicle charging
  infrastructure assessment: Analyzing charging needs to support zero-emission
  vehicles in 2030 – revised staff report.}
\newblock \bibinfo{type}{Tech. Rep.} \bibinfo{number}{CEC-600-2021-001-REV},
  \bibinfo{institution}{California Energy Commission} (\bibinfo{year}{May
  2021}).

\bibitem{montenegro2012real}
\bibinfo{author}{Montenegro, D.}, \bibinfo{author}{Hernandez, M.} \&
  \bibinfo{author}{Ramos, G.}
\newblock \bibinfo{title}{Real time opendss framework for distribution systems
  simulation and analysis}.
\newblock \emph{\bibinfo{journal}{Sixth IEEE/PES Transmission and Distribution:
  Latin America Conference and Exposition (T\&D-LA)}} \bibinfo{pages}{1--5}
  (\bibinfo{year}{2012}) .

\bibitem{CEC_appliance}
\bibinfo{author}{CEC}.
\newblock \bibinfo{title}{Modernized appliance efficiency database system
  (maedbs)}.
\newblock
  \bibinfo{howpublished}{\url{https://cacertappliances.energy.ca.gov/Pages/ApplianceSearch.aspx}}.
\newblock \bibinfo{note}{Accessed: 2023}.

\bibitem{NREL_EV}
\bibinfo{author}{Muratori, M.}, \bibinfo{author}{Rames, C.~L.},
  \bibinfo{author}{Srinivasa~Raghavan, S.}, \bibinfo{author}{Melaina, M.~W.} \&
  \bibinfo{author}{Wood, E.~W.}
\newblock \bibinfo{title}{National plug-in electric vehicle infrastructure
  analysis}  (\bibinfo{year}{2018}).
\newblock \urlprefix\url{https://www.osti.gov/biblio/1420371} .

\bibitem{storage_efficiency}
\bibinfo{author}{Mongird, K.} \emph{et~al.}
\newblock \bibinfo{title}{2020 grid energy storage technology cost and
  performance assessment}.
\newblock \bibinfo{type}{Tech. Rep.}, \bibinfo{institution}{USDOE}
  (\bibinfo{year}{December 2020}).
\newblock
  \urlprefix\url{https://www.energy.gov/energy-storage-grand-challenge/downloads/2020-grid-energy-storage-technology-cost-/
  and-performance}.

\bibitem{NREL_solar}
\bibinfo{author}{Gagnon, P.}, \bibinfo{author}{Margolis, R.},
  \bibinfo{author}{Melius, J.}, \bibinfo{author}{Phillips, C.} \&
  \bibinfo{author}{Elmore, R.}
\newblock \bibinfo{title}{Rooftop solar photovoltaic technical potential in the
  united states: A detailed assessment}.
\newblock \bibinfo{type}{Tech. Rep.} \bibinfo{number}{NREL/TP-6A20-65298},
  \bibinfo{institution}{NREL}, \bibinfo{address}{Golden, CO, USA}
  (\bibinfo{year}{2016}).
\newblock \urlprefix\url{https://www.nrel.gov/docs/fy16osti/65298.pdf}.

\bibitem{NREL_solar_data}
\bibinfo{author}{NREL}.
\newblock \bibinfo{title}{Solar power data for integration studies}.
\newblock
  \bibinfo{howpublished}{\url{https://www.nrel.gov/grid/solar-power-data.html}}.
\newblock \bibinfo{note}{Accessed: 2023}.

\bibitem{EV_charging_behaviors}
\bibinfo{author}{Chakraborty, D.}, \bibinfo{author}{Bunch, D.~S.},
  \bibinfo{author}{Lee, J.~H.} \& \bibinfo{author}{Tal, G.}
\newblock \bibinfo{title}{Demand drivers for charging infrastructure-charging
  behavior of plug-in electric vehicle commuters}.
\newblock \emph{\bibinfo{journal}{Transportation Research Part D: Transport and
  Environment}} \textbf{\bibinfo{volume}{76}}, \bibinfo{pages}{255--272}
  (\bibinfo{year}{2019}).
\newblock
  \urlprefix\url{https://www.sciencedirect.com/science/article/pii/S1361920919301919}.
\newblock \doi{https://doi.org/10.1016/j.trd.2019.09.015} .

\bibitem{EV_model_siobhan}
\bibinfo{author}{Powell, S.}, \bibinfo{author}{Cezar, G.~V.} \&
  \bibinfo{author}{Rajagopal, R.}
\newblock \bibinfo{title}{Speech original model}.
\newblock \emph{\bibinfo{journal}{Mendeley Data}} \textbf{\bibinfo{volume}{V1}}
  (\bibinfo{year}{2021}).
\newblock \doi{10.17632/gvk34mybtb.1} .

\bibitem{PF_linear_survey}
\bibinfo{author}{Bolognani, S.} \& \bibinfo{author}{Dörfler, F.}
\newblock \bibinfo{title}{Fast power system analysis via implicit linearization
  of the power flow manifold}.
\newblock \emph{\bibinfo{journal}{53rd Annual Allerton Conference on
  Communication, Control, and Computing (Allerton)}} \bibinfo{pages}{402--409}
  (\bibinfo{year}{2015}).
\newblock \doi{10.1109/ALLERTON.2015.7447032} .

\bibitem{LinearPF_LSQ}
\bibinfo{author}{Liu, Y.}, \bibinfo{author}{Zhang, N.}, \bibinfo{author}{Wang,
  Y.}, \bibinfo{author}{Yang, J.} \& \bibinfo{author}{Kang, C.}
\newblock \bibinfo{title}{Data-driven power flow linearization: A regression
  approach}.
\newblock \emph{\bibinfo{journal}{IEEE Transactions on Smart Grid}}
  \textbf{\bibinfo{volume}{10}}, \bibinfo{pages}{2569--2580}
  (\bibinfo{year}{2019}) .

\end{thebibliography}



\end{document}